\documentclass{article}
\usepackage[utf8]{inputenc}
\usepackage[a4paper, total={6.5in, 9.2in}]{geometry}
\usepackage{authblk}
\pdfoutput=1

\title{Node-reconfiguring multilayer networks of human brain function}
\author[1,*]{Tarmo Nurmi}
\author[1]{Pietro De Luca}
\author[2,3]{Maria Hakonen}
\author[1]{Mikko Kivel\"a}
\author[1,4]{Onerva Korhonen}
\affil[1]{Department of Computer Science, Aalto University School of Science, P.O. Box 15400, Aalto FI-00076, Finland}
\affil[2]{Athinoula A. Martinos Center for Biomedical Imaging, Department of Radiology, Massachusetts General Hospital Charlestown, Boston, MA, USA}
\affil[3]{Department of Radiology, Harvard Medical School, Boston, MA, USA}
\affil[4]{Faculty of Science, Forestry and Technology, University of Eastern Finland, Joensuu campus, P.O. Box 111, Joensuu FI-80101, Finland}
\affil[*]{To whom correspondence should be addressed: tarmo.nurmi@aalto.fi}
\date{}

\usepackage{graphicx}
\usepackage{subcaption}
\usepackage{amssymb}
\usepackage{hyperref}
\usepackage{mwe}
\usepackage{xcolor}
\usepackage{xr}
\usepackage{multirow}
\usepackage{booktabs}
\usepackage{enumitem}

\newcommand\abs[1]{\left|#1\right|}

\emergencystretch=\maxdimen
\begin{document}
\maketitle

\begin{abstract}

The properties of functional brain networks are heavily influenced by how the network nodes are defined. A common approach uses Regions of Interest (ROIs), which are predetermined collections of functional magnetic resonance imaging (fMRI) measurement voxels, as network nodes. Their definition is always a compromise, as static ROIs cannot capture the dynamics and the temporal reconfigurations of the brain areas. 
Consequently, the ROIs do not align with the functionally homogeneous regions, which can explain the very low functional homogeneity values observed for the ROIs. This is in violation of the underlying homogeneity assumption in functional brain network analysis pipelines and it can cause serious problems such as spurious network structure.
We introduce the node-reconfiguring multilayer network model, where nodes represent ROIs with boundaries optimized for high functional homogeneity in each time window. In this representation, network layers correspond to time windows, intralayer links depict functional connectivity between ROIs, and interlayer link weights quantify the overlap between ROIs on different layers.
The ROI optimization approach increases functional homogeneity notably, yielding an over 10-fold increase in the fraction of ROIs with high homogeneity compared to static ROIs from the Brainnetome atlas.  
The optimized ROIs reorganize non-trivially at short time scales of consecutive time windows and across several windows. The amount of reorganization across time windows is connected to intralayer hubness: ROIs that show intermediate levels of reorganization have stronger intralayer links than extremely stable or unstable ROIs.
Our results demonstrate that reconfiguring parcellations yield more accurate network models of brain function. This supports the ongoing paradigm shift towards the chronnectome that sees the brain as a set of sources with continuously reconfiguring spatial and connectivity profiles.
\end{abstract}


\textbf{Key points:}
\begin{itemize}[topsep=0pt, noitemsep]
    \item We introduce the node-reconfiguring multilayer network model, a functional brain network model with nodes that change at short time scales.
    \item Our optimization approach notably increases the functional homogeneity of ROIs compared to the static ROIs from the Brainnetome atlas.
    \item Our model catches non-trivial short-term reconfiguration of ROIs and opens access to various multilayer network analysis methods previously untapped in network neuroscience.
\end{itemize}

\section{Introduction}

Network neuroscience \cite{muldoon2016network, bassett2017network} models the brain as a complex network. The nodes of functional brain networks depict brain areas, while edges represent temporal co-activation of the areas \cite{korhonen2021principles}. Observed properties of functional brain networks depend on the methodological choices made in preprocessing \cite{andellini2015test, aurich2015evaluating} and network construction \cite{fornito2013graph, korhonen2021principles}. In particular, definition of network nodes strongly affects network properties \cite{wang2009parcellation, zalesky2010whole}. 

Despite the importance of accurate node definition for obtaining reliable analysis outcomes, network neuroscience still lacks a commonly shared standard for node definition \cite{zalesky2010whole, eickhoff2015connectivity, korhonen2021principles}. The most popular approach uses as nodes Regions of Interest (ROIs), collections of functional magnetic resonance imaging (fMRI) measurement voxels defined in terms of, e.g., anatomy, function, or connectivity \cite{korhonen2021principles}. The ROI time series are obtained by averaging the fMRI BOLD time series of voxels inside each ROI. Unfortunately, the ROI approach suffers from two important drawbacks.

\begin{figure}
    \centering
    \includegraphics[width=0.8\textwidth]{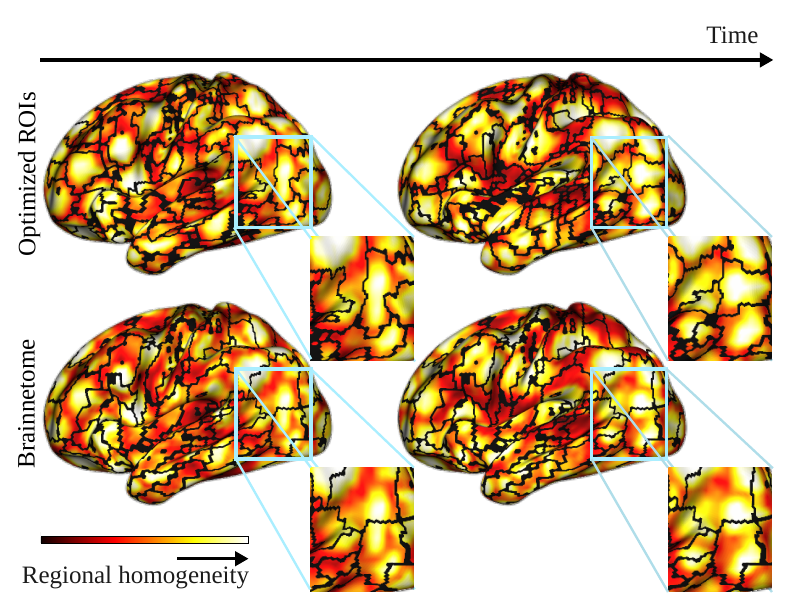}
    \caption{Optimized ROIs are necessary for modelling time-dependent functional connectivity. Coloring of the brain surface shows Regional Homogeneity (ReHo, Equation \ref{eq:kendalls_tau}) calculated from simulated data with homogeneous random ROIs in two time windows. Optimized ROI boundaries (top row) match the groups of similarly behaving voxels reflected by high ReHo values, while static parcellations like Brainnetome (bottom row) break these groups, yielding low functional homogeneity of ROIs and data losses. To simulate the data, we created 210 random ROIs covering the cerebral cortex as described in section \ref{sec:random_rois}, assigned each ROI seed with a white noise time series, and defined the time series of other voxels as a mixture $y(t) = G(d)/G(0)x(t) + (1 - G(d)/G(0))n$ where $G$ is the Gaussian function with mean 0 and variance 5, $d$ is the Euclidian distance between the voxel and the seed of its ROI, $x(t)$ is the time series of the ROI seed, and $n$ is a white noise time series. Optimized ROI boundaries were obtained using the greedy algorithm with weighted mean consistency as priority function. For easier visualization, the data was simulated and the optimized ROIs obtained in the MNI standard space, while all the remaining analysis of this article were performed in each subject's native space.}
    \label{fig:brain_surface_parcellation_visualization}
\end{figure}

First, using ROIs as functional brain network nodes requires functional homogeneity, or high correlation between voxel time series inside each ROI. fMRI measurement data is assumed to consists of groups of highly correlated voxels (represented with bright colors in Figure~\ref{fig:brain_surface_parcellation_visualization}). To avoid losing data when averaging voxel time series, ROIs should match these voxel groups, as the two parcellations on the top row of Figure~\ref{fig:brain_surface_parcellation_visualization} do. Unfortunately, most commonly used parcellations resemble more those on the bottom row of Figure~\ref{fig:brain_surface_parcellation_visualization}, where ROIs do not match the groups of correlated voxels \cite{gottlich2013altered, stanley2013defining, korhonen2017consistency}. In other words, many commonly used ROIs suffer from low functional homogeneity. The low voxel-level correlations yield data losses when voxel time series are averaged \cite{stanley2013defining}. Further, investigation of connectivity at the levels of ROIs and voxels has revealed that many ROI pairs are strongly connected even though the voxels of the same ROIs show only low connectivity \cite{korhonen2017consistency}. This suggests that low homogeneity may yield spuriosities in observed ROI-level network structure, for example because badly defined ROI boundaries divide groups of correlated voxels into several ROIs.

{}

Second, a vast majority of ROI definition approaches produce static \textit{a priori} parcellations of the brain, that is, ROIs with fixed boundaries that do not vary between subjects or over time. This is in line with the current functional network construction paradigm: functional brain networks are often constructed over whole measurement time series, ignoring time-dependent changes in connectivity \cite{hutchison2013dynamic, ryyppo2018regions, iraji2020space, 
calhoun2014chronnectome}. However, more detailed time-window-level analyses have revealed that voxel-level connectivity inside ROIs fluctuates in time and that these fluctuations probably cause the observed low functional homogeneity values of ROIs when whole measurement time series are used \cite{ryyppo2018regions, luo2021within}. Further, studies that cluster voxel time series separately inside each time window to create maximally homogeneous ROIs have shown that optimal ROIs differ between subjects and reorganize even within a single subject, both between cognitive tasks and spontaneously in rest \cite{iraji2019spatialdynamics, salehi2020there}. The time (horizontal) dimension of Figure \ref{fig:brain_surface_parcellation_visualization} demonstrates the challenge that the time-dependent connectivity variations present to ROI definition: when the correlated voxel groups change in time, ROIs can match them only if ROI boundaries are also allowed to fluctuate. The first application of time-dependent nodes in brain network analysis \cite{luo2021within} has shown that time-dependent ROIs reduce spuriosities in observed network structure and therefore open new insights on the connectivity differences between subjects and tasks.

Here, we propose the node-reconfiguring multilayer network model of brain function.
In our model, the nodes are optimized ROIs that fluctuate over short time scales. The model builds on the theory of multilayer networks. These networks combine different types of connections into a single network so that connections of each type form the intralayer edges on the layer of their own, and interlayer edges connect the different layers \cite{kivela2014multilayer, boccaletti2014structure}. Multilayer networks have established their place in the toolbox of network neuroscience \cite{de2017multilayer, vaiana2018multilayer}, for example in the study of the connection between brain structure and function \cite{battiston2017multilayer, crofts2016structure} and of nested functional connectivity at different MEG and EEG frequency bands \cite{brookes2016multi, guillon2017loss, yu2017selective}. The layers of our model correspond to time windows similarly as in, e.g., \cite{bassett2011dynamic, bassett2013task, braun2015dynamic, telesford2017cohesive}. Unlike these studies that used static ROIs as nodes, we optimize the nodes separately for each subject and time window to maximize functional homogeneity.
To this end, we use a greedy algorithm inspired by the work of Salehi et al. \cite{salehi2018exemplar} and the Craddock algorithm \cite{craddock2012whole}. Intralayer edges represent the functional connectivity in each time window, while interlayer edges depict the spatial overlap between ROIs on different layers.

Our ROI optimization approach increased functional homogeneity of ROIs significantly. Together with the fluctuations we observe in ROI boundaries between layers, this confirms that the boundaries of correlated voxel groups change at short time scales of $\sim$10 s. Through rigorous comparisons, we show the superiority of the node-reconfiguring model compared to more simple models, in particular two variants of multiplex networks (i.e., multilayer networks that contain only diagonal interlayer edges that connect the instances of the same nodes on different layers) constructed using static ROIs as nodes. Therefore, the more accurate fit of our model to the functional neuroimaging data justifies the increased complexity of the model.

Our results support the ongoing paradigm shift in network neuroscience. Until recently, network neuroscientists have considered functional connectivity static or, at most, investigated changing functional edges between static nodes. Our results, however, are in line with the new chronnectome paradigm \cite{calhoun2014chronnectome, iraji2019spatialchronnectome, iraji2020space} that sees the brain as a collection of sources with time-dependent spatial and connectivity profiles. Our ROI optimization methods and multilayer network construction approach are available as a public Python implementation, which allows the network neuroscience community to adopt these tools and build on top of them. Importantly, the modular structure of our implementation allows for replacing the ROI optimization methods used in the present study with any time series clustering approach. Further, our model does not only produce a more accurate network representation of the brain but also opens the possibility for further exploring brain function with several analysis methods tailored for multilayer networks, for example, multilayer clustering and motif analysis.

\section{Materials and methods}

\subsection{Subjects}

We used fMRI data from 25 subjects (13 female) freely listening to an audiobook. The data was originally collected for a study on the effects of cultural background on speech and narrative processing \cite{hakonen2022processing}; here, we used the Finnish subject group. The age of the subjects varied between 19 and 35 years (mean age 25.8$\pm$4.45 years, mean$\pm$std). None of the subjects reported any neurological or hearing deficits. All subjects gave their written, informed consent before participating in the data collection. The study was approved by the Aalto University Research Ethics Committee and followed the principles of the Declaration of Helsinki.

\subsection{Data acquisition}

The data was collected with a 3T MRI whole-body scanner (MAGNETOM Skyra, Siemens Healthcare, Erlangen, Germany) using a 32-channel receiving head coil array. The anatomical MR images were acquired with a T1-weighted MPRAGE sequence (TR = 2530 ms, TE = 3.30 ms, FOV = 256 mm, flip angle = 7 degrees, slice thickness = 1 mm). fMR images were acquired with an ultrafast simultaneous multislice inverse imaging (InI) sequence \cite{hsu2017simultaneous} (TE = 27.5 ms, flip angle = 30°, FOV = 210 × 210 × 210 mm3, in-plane resolution = 5 mm × 5 m) in 9 runs with duration ranging from 4.8 to 8.4 min. The InI encoding direction was superior-inferior, while spatial information in anterior-posterior and left-right directions was recovered with frequency and phase encoding. First, 24 axial slices of 7 mm were collected without inter-slice gap. These slices were divided into two equally-sized groups, and both groups were excited and read in 50 ms, which yielded a TR of 100 ms. Adjacent slices in both groups were separated using simultaneous echo refocusing \cite{feinberg2013ultra}, and blipped-controlled aliasing in parallel imaging \cite{setsompop2012blipped} was used for further aliasing control. For full details of data acquisition, see \cite{hakonen2022processing}.

\subsection{Preprocessing}

The anatomical images were reconstructed with Freesurfer's recon-all tool (\url{http://surfer.nmr.mgh.harvard.edu/}) and the functional images with the regularized sensitivity encoding (SENSE) tool \cite{lin2004parallel, lin2005functional} (regularization parameter 0.005). First 12.3 s of each fMRI measurement were removed, scanner drift was removed with a Savitzky–Golay filter \cite{ccukur2013attention} (order: 3, frame length: 240 s), and physiological and movement-related noise was removed with the MaxCorr method \cite{pamilo2015correlation}. Data was zero-phase filtered between 0.08 Hz and 4 Hz. Unlike in \cite{hakonen2022processing}, no spatial smoothing was applied because of its potentially biasing effects on network analysis \cite{alakorkko2017effects}. Further, to obtain nodes that optimally match the individual anatomy of each subject, we did not register the data to any template but defined nodes (see section \ref{sec:node-definition}) and performed all analysis in each subject's native space.

\subsection{Definition of network nodes}
\label{sec:node-definition}

The nodes of our node-reconfiguring multilayer network model correspond to ROIs defined individually for each subject and time window. ROI definition has two goals. First, we search for ROIs with \textit{high functional homogeneity}, defined in terms of spatial consistency \cite{korhonen2017consistency}:

\begin{equation} \label{eq:consistency}
    \phi(I)=\frac{1}{|I|(|I|-1)}\sum_{u, v\in I; u \neq v} C(u,v),
\end{equation}
where $|I|$ depicts the size of ROI $I$, defined as the number of voxels in the ROI, the summation goes over all (ordered) pairs of voxels $u,v$ in the ROI, and $C(u,v)=C(x_v(t),x_{u}(t))$ is the Pearson correlation coefficient between the time series of voxels $v$ and $u$, $x_v(t)$ and $x_u(t)$. In other words, $\phi(I)$ is the mean correlation of voxel time series in $I$. Second, the \textit{size distribution of ROIs should be relatively narrow} with no extremely large or small ROIs. In particular, we want to keep down the number of ROIs that contain only a single voxel, since single-voxel ROIs are probably too small to match real functional brain areas and thus reflect rather noise than true underlying brain function.

To define the ROIs, we used two variants of the greedy clustering introduced by Salehi et al. in~\cite{salehi2018exemplar} and the Craddock, or NCUT spectral clustering, algorithm \cite{craddock2012whole}, and as reference methods the static Brainnetome atlas \cite{fan2016human} and random ROIs. We aimed to define 246 ROIs with each approach; this number was selected to match the number of ROIs in the Brainnetome atlas defined in the MNI standard space. However, the number of ROIs varied slightly between subjects due to details of the clustering algorithms. The native-space transformation can discard some Brainnetome ROIs, yielding a slightly lower average number of ROIs across subjects, runs, and time windows in native space. The greedy clustering algorithms use seeds based on native-space Brainnetome ROIs and thus inherit the number of ROIs from them. The Craddock approach is known to produce less ROIs than desired \cite{craddock2012whole}. For details on the number and size of ROIs, see Table \ref{table:parcellation_props} in section \ref{sec:optimization_increases_homogeneity}.

\subsubsection{Greedy clustering algorithm (weighted mean consistency and minimum correlation)}

The greedy clustering algorithm for defining optimized ROIs begins by selecting ROI seeds. Then, the algorithm initializes a priority queue of pairs of all ROIs and all their neighboring voxels (i.e. voxels which share a cuboid face with a voxel in the ROI) that have not yet been assigned to any ROI. Therefore, a voxel can appear in the priority queue multiple times paired with different ROIs. A priority value is calculated for each ROI-voxel pair, and at each iteration of the algorithm, the voxel of the ROI-voxel pair with the highest priority value is selected as the candidate voxel. The candidate voxel can be directly added to the corresponding ROI, after which the priority queue and priority values are updated, possibly adding new ROI-voxel pairs to the queue. However, applying a threshold at this stage further improves the functional homogeneity of ROIs; in this case, the candidate voxel is added to the corresponding ROI only if its priority value surpasses a given threshold. The algorithm stops when all voxels in the queue are either assigned to an ROI (the queue is empty) or excluded by thresholding. Since the ROIs are grown one spatially contiguous voxel at a time, the final ROIs are also spatially contiguous. This is a specifically engineered feature of the clustering: to represent functional units that ultimately consist of neurons, ROIs must be spatially contiguous to ensure that there is a possibility for physical connection between neurons inside each ROI.

Due to the modular nature of the algorithm, seed selection strategy, priority value definition, and thresholding approach can be selected independently. We selected the seeds based on regional homogeneity (ReHo) \cite{zang2004regional}, defined as the Kendall's coefficient of concordance \cite{kendall1990rank} among the voxel's neighbors:

\begin{equation} \label{eq:kendalls_tau}
    W = \frac{12\sum_{i=1}^n (R_{i} - \overline{R})^2}{K^2(n^3-n)},
\end{equation}
where the summation goes over time, $n$ being the total number of time points, and $K$ depicts the number of time series used for calculating $W$. Here, we defined neighborness as sharing a face with the central voxel, which yielded $K = 7$ (the 6 neighbors and the central voxel itself). The rank sum $R_i$ at the $i$th time point is defined as
\begin{equation} \label{eq:rank_sum}
    R_i = \sum_{j=1}^Kr_{i,j},
\end{equation}
where the summation goes across the voxels in the neighborhood and $r_{i,j}$ is the rank of the $i$th value in the time series of voxel $j$. $\overline{R}$ is the mean of $R_i$ across time.

We calculated the ReHo of each voxel separately in each time window, selected from each ROI of the native-space Brainnetome atlas (see section \ref{section:brainnetome}) the voxel with the highest ReHo and used these voxels as seeds for this window. This seed selection strategy ensured that the seeds are located in a homogeneous neighbourhood and relatively equally distributed in the brain space and that the number of ROIs matches the number of the Brainnetome ROIs used as a reference.

To define the priority values of the ROI-voxel pairs $I,i$, we used two alternative measures. In the \emph{weighted mean consistency} approach, the priority value was defined as:

\begin{equation} \label{eq:weighted_consistency}
    PV(I,i) = \frac{\sum_{I' \in ROIs}|I'|\phi(I')}{\sum_{I' \in ROIs}|I'|} - \lambda\frac{\sum_{I' \in ROIs}|I'|^2}{(\sum_{I' \in ROIs}|I'|)^2},
\end{equation}
where spatial consistency $\phi(I)$ and size $|I|$ are calculated assuming that the candidate voxel $i$ has been assigned to ROI $I$, and $\lambda$ is a regularization parameter. The first term of Eq. \ref{eq:weighted_consistency} maximizes functional homogeneity of ROIs, while the second term is used for regularization to control the width of the size distribution. For the regularization, we used $\lambda$=100; for details on the selection of this value, see section \ref{sec:SI_selection_of_params} in the Supplementary Information. 

Alternatively, in the \emph{minimum correlation} (min correlation) approach, the priority value $PV(I,i)$ was defined as the lowest Pearson correlation coefficient between the time series of the candidate voxel and the time series of any voxel already assigned to ROI $I$. The rationale behind this approach is to simplify the priority function without compromising the size distribution. The weighted mean consistency approach requires regularization, as maximizing \emph{mean} spatial consistency may increase the number of small ROIs that tend to have higher consistency \cite{korhonen2017consistency}. On the other hand, maximizing the number of high between-voxel correlations by controlling the \emph{lowest} correlation within each ROI does not affect size distribution and therefore does not warrant a regularization term.

At the thresholding step, we calculated the average Pearson correlation coefficient between the time series of the candidate voxel and all voxels of each ROI; this equals to the spatial consistency of each ROI assuming that the candidate voxel has been assigned to this ROI. The candidate voxel was added to its corresponding ROI only if the average correlation between the candidate voxel and the ROI was among the $N\%$ highest ones. Here, we used $N$=30; for details on the selection of the threshold value, see section \ref{sec:SI_selection_of_params}.

It is possible that not all seed voxels of the greedy clustering algorithm are located in a particularly homogeneous neighbourhood, even if a seed selection strategy based on ReHo is used. Stringent enough thresholding may prevent these seeds from growing in the first place, yielding ROIs that consists of a single voxels. These ROIs are most probably an artifact caused by the predefined number of seeds. Therefore, we excluded these single-voxel ROIs from all further analysis, and only reported their number for all investigated parcellations (Table \ref{table:parcellation_props}).

\subsubsection{Craddock algorithm (normalized cut spectral clustering)}

The Craddock clustering begins by constructing a weighted network of voxels. In this network, each voxel is connected to its six neighbors, and edge weights are defined as the Pearson correlation coefficient between voxel time series. The network is then thresholded by removing edges that have a weight lower than a fixed threshold. Here, we used threshold 0.2; for details on the selection of the threshold value, see section \ref{sec:SI_selection_of_params}.

Then, the network is divided into clusters in a way that minimizes the NCUT cost

\begin{equation} \label{eq:ncut}
    \textrm{NCUT}(A,B) = \frac{\textrm{CUT}(A,B)}{\sum_{i\in A, n \in V}w_{i,n}} + \frac{\textrm{CUT}(A,B)}{\sum_{j \in B, n \in V}w_{j,n}},
\end{equation}
where $A$ and $B$ are sets of voxels in two clusters to be separated by cutting edges, $V$ is the set of all voxels in the network, $w_{i,j}$ is the link weight between voxels $i$ and $j$, and CUT$(A,B)$ is defined as

\begin{equation} \label{eq:cut}
    \textrm{CUT}(A,B) = \sum_{i \in A, j \in B}w_{i,j}.
\end{equation}
The desired number of clusters is given as a free parameter; here we used the number of ROIs in the Brainnetome atlas, 246.

Note that the Craddock algorithm and Craddock parcellations used in this article are different from the commonly used Craddock atlas; the later is an example outcome of the algorithm, also published in \cite{craddock2012whole} and commonly used as a static \textit{a priori} parcellation.

\subsubsection{Brainnetome atlas}
\label{section:brainnetome}

The Brainnetome atlas is a multimodal ROI atlas based on brain anatomy and structural and functional connectivity. Brainnetome contains 246 ROIs, of which 210 cover the cerebral cortex and 36 the subcortical gray matter. Before network analysis, we transformed the Brainnetome template to each subject's native space with the SPM 12 software (\url{https://www.fil.ion.ucl.ac.uk/spm/software/spm12/}). Note that the transformation in some cases lead to different numbers of ROIs in different subjects; for details see Table \ref{table:parcellation_props}.

\subsubsection{Random ROIs}
\label{sec:random_rois}

To construct random ROIs, we first selected 246 random seed voxels per time window. Then, at each iteration, we identified for each ROI the neighboring voxels that had not yet been assigned to any ROI and assigned them to this ROI. The process continued until all voxels had been assigned to an ROI and produced approximately spherical ROIs.

\subsection{Multilayer network construction}

\begin{figure}
    \centering
    \includegraphics[width=0.7\textwidth]{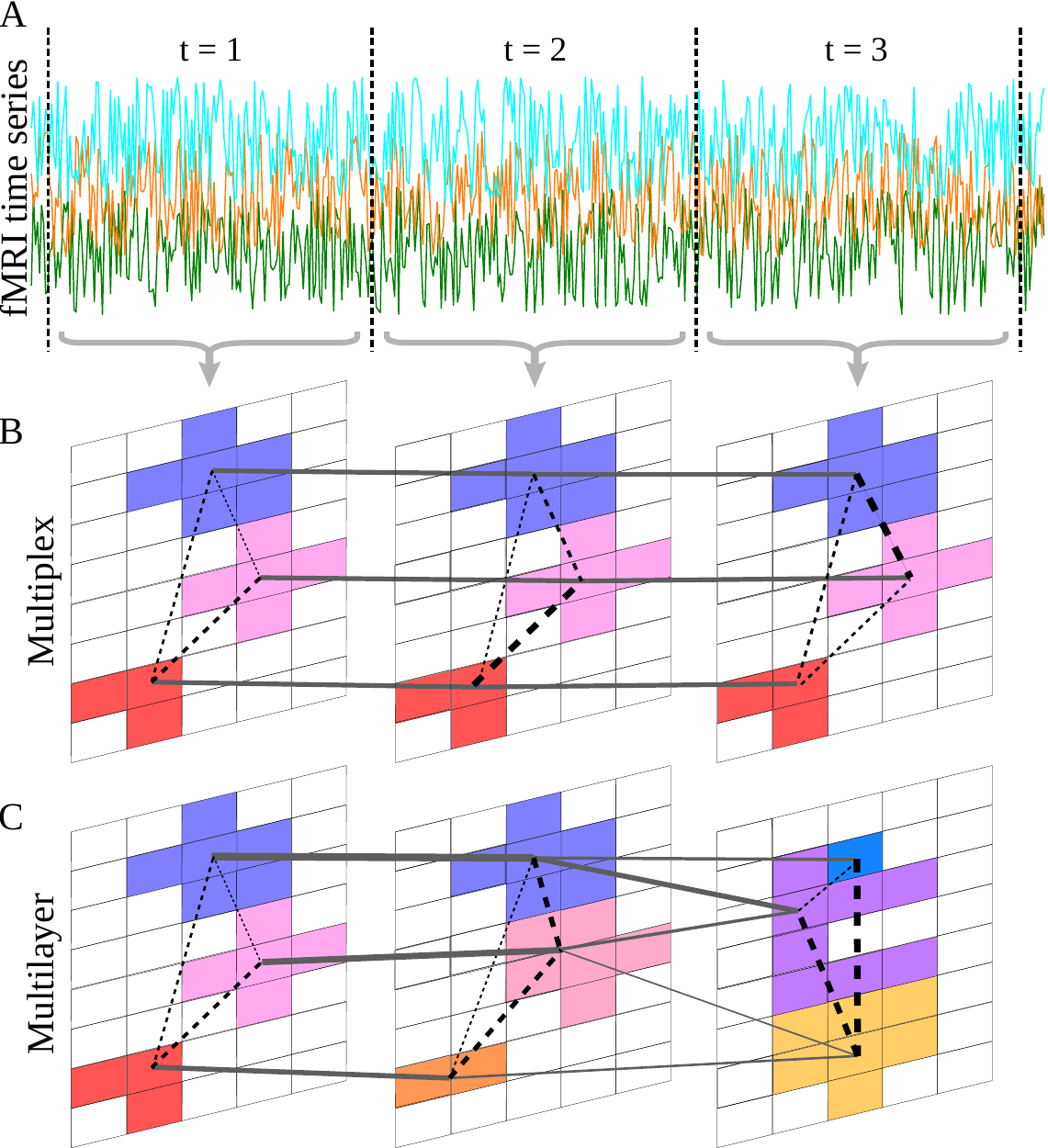}
    \caption{Schematic illustration of the node-reconfiguring multilayer network construction approach. A) The fMRI time series are divided into time windows (t = 1, 2, and 3) that are represented by the network layers. B) In the multiplex network, the same nodes (Brainnetome ROIs) are used on each layer; here, the background grid corresponds to voxels, and the colored areas represent the nodes. Edges inside layers (dashed lines) represent the functional connectivity (similarity of time series) between the nodes, while interlayer edges have unit weight and connect the instances of the same node on consecutive layers. C) In the node-reconfiguring multilayer network, nodes are defined separately for each time window as contiguous sets of voxels clustered using data from that specific time window. Depending on the node-definition approach, some voxels may remain outside ROIs due to thresholding. Note that nodes do not have a stable identity across layers, and may also disappear and (re-)appear from layer to layer. Intralayer edges represent functional connectivity, and interlayer edges are weighted according to the Jaccard index of the sets of voxels of ROIs on consecutive layers, representing spatial overlap and thus ROI fluctuation.}
    \label{fig:schema}
\end{figure}

We constructed a multilayer network for each subject and imaging run. We first divided the data into non-overlapping time windows that formed the layers of the network (Figure \ref{fig:schema}A). The window length, 80 samples, was selected based on earlier research \cite{ryyppo2018regions} to ensure stable values of spatial consistency. The duration of the window in real time was 8 seconds (80 samples $\times$ 0.1 seconds/sample). Then, we applied the ROI definition methods described above separately in each window, to create the set of nodes (ROIs) for each layer (visualized with different colours in Figure~\ref{fig:schema}B-C).

We obtained the time series of each ROI by averaging the time series of the voxels in the ROI. The weight of intralayer edges depicted the Pearson correlation coefficient between ROI time series within each time window.

To define the interlayer edge weights, we calculated the spatial overlap (i.e. fraction of shared voxels) between each pair of ROIs on temporally consecutive layers. The weight of an interlayer edge reflected the change an ROI undergoes from one time window to the next. More specifically, let $I_t$ be the set of voxels of an ROI on layer $t$, and let $I'_{t+1}$ be the set of voxels of an ROI on layer $t+1$. The weight of the interlayer edge between the ROIs was set to be the Jaccard index of the voxel sets, that is, the cardinality of their intersection divided by the cardinality of their union $\frac{\abs{I_t \cap I'_{t+1}}}{\abs{I_t \cup I'_{t+1}}}$. If the ROIs are the same, the weight is 1, and if they do not share any voxels, the weight is 0 (the edge does not exist). 

In the networks constructed using Brainnetome ROIs, all interlayer edges have unit weight due to the static nature of the atlas (Figure~\ref{fig:schema}B). In other words, these networks are multiplex: all interlayer edges connect ROIs and their identical counterparts on the consecutive layers.

In the networks constructed with the other node definition approaches (that is, in node-reconfiguring multilayer networks), an ROI on layer $t$ could connect to more than one ROI on the subsequent layer $t+1$, and an ROI on the subsequent layer $t+1$ could have connections to more than one ROI on the previous layer $t$. This yielded multilayer networks (Figure \ref{fig:schema}C), where nodes from one layer can connect to any other nodes on the subsequent layer (in contrast to multiplex networks, where a node is only connected to one single counterpart node on the subsequent layer). The mathematical framework for representing such general multilayer networks is described by Kivelä et al. in \cite{kivela2014multilayer}.

{}

\subsection{Network analysis}
\label{section:network_analysis}

We examined 1) the stability and instability behavior of ROIs across time, and 2) interactions of fundamental intra- and interlayer properties of the networks.

\subsubsection{Stability behavior of ROIs across time}

We opened the analysis of ROIs' stability across time with \emph{interlayer edge weight} distributions. Then, we defined ROI \emph{stability score} which measures ROI's stability from one layer (time window) to the next. For ROI $I$ on layer $t$, the stability score is the weight of the highest-weight interlayer edge connecting $I$ to any ROI on layer $t+1$. In other words, the stability score is the highest overlap of an ROI with another ROI on the next layer. If the ROI does not change at all, its stability score is 1. 

Changes in ROI boundaries affect the stability score in three ways; let us consider ROI $I$ with size $|I|$ on layer $t$ as an example. First, some of the voxels assigned to ROI $I$ on layer $t$ can be assigned to some other ROI on layer $t+1$. For example, if 70 \% of the voxels belonging to ROI $I$ on layer $t$ remain together to form an ROI $I^*$ on layer $t+1$ (while 30 \% of the voxels of $I$ are assigned to other ROIs), then the size of the intersection of $I$ and $I^*$ is 0.7$|I|$ and their union is $|I|$ (as all voxels of $I^*$ belong to $I$ on layer $t$). So, the Jaccard index between $I$ and $I^*$ and thus $I$'s stability score is $\frac{0.7|I|}{|I|}$ = 0.7.
Second, some voxels assigned to ROIs other than $I$ on layer $t$ can be included in ROI $I^*$ on layer $t+1$. For example, if all voxels of ROI $I$ remain together and an additional 30 \% more come from other ROIs (that is, the total size of $I^*$ is 1.3$|I|$), the size of the intersection of $I$ and $I^*$ is $|I|$, their union is now 1.3$|I|$ (as all voxels of $I$ belong also to $I^*$), and the stability score of $I$ is thus $\frac{|I|}{1.3|I|} \approx 0.77$.
Finally, removal and merging of voxels can, and often do, happen at the same time. For example, if 30 \% of $I$'s voxels are assigned to other ROIs on layer $t+1$ but at the same time an additional 70 \% (of $|I|$) of voxels assigned to other ROIs at $t$ are merged to $I^*$ (so that the size of $I^*$ is now $2\times 0.7|I|$), the intersection of $I$ and $I^*$ remains as $0.7|I|$, their union is $0.3|I| + 0.7|I| + 0.7|I|$ (that is, voxels that belong only to $I$, voxels that belong to both $I$ and $I^*$, and voxels that belong only to $I^*$), and the stability score of $I$ is $\frac{0.7|I|}{(0.3+0.7+0.7)|I|} \approx 0.41$. 

To further investigate ROIs' stability and behavior in time, we looked at the individual evolution of ROIs across layers. We defined the \emph{trajectory} of an ROI on layer $t$ as the time series of highest Jaccard indexes between the voxel sets of that ROI and ROIs on temporally subsequent layers. In other words, for an ROI on layer $t$, the first number of the trajectory is calculated by finding the ROI on layer $t+1$ which has the highest Jaccard index with the original ROI and setting the value of the trajectory to that Jaccard index. The second number is the highest Jaccard index between the original ROI on layer $t$ and ROIs on layer $t+2$, and so on.

The trajectory describes the reconfiguration behavior of the original ROI across time. Similarly to the stability score, if the ROI does not change, then the trajectory has value 1. If 70 \% of the voxels in an ROI remain together in an ROI on a subsequent layer, and no other voxels are added, then the trajectory has a value of 0.7. If all the voxels are excluded from ROIs due to thresholding on a subsequent layer, the ROI disappears completely and the trajectory has a value of 0. The trajectory of an ROI can point out if an ROI returns (close to) the original configuration after being split or shattered.

\subsubsection{Interactions of intra- and interlayer network properties}

We investigated the interplay of intra- and interlayer network properties by analyzing two intralayer network characteristics, \emph{mean intralayer correlation} and \emph{ROI edge weight}, in relation to stability score and \emph{pairwise stability score}. Mean intralayer correlation is an ROI-specific property defined as the mean intralayer edge weight of an ROI within the layer where the ROI resides. Mean intralayer correlation reflects the connectivity of the ROI with other ROIs in a single time window. To put the connectivity of an ROI into context with its stability, we looked at mean intralayer correlation as a function of the stability score. 

ROI edge weight is an ROI-pair-specific property, equal to the intralayer edge weight between two ROIs on the same layer. It describes the functional connectivity of two ROIs within a time window. To contextualize ROI edge weight with the interlayer stabilities of the ROIs, we looked at it as a function of pairwise stability score, defined as the mean of the stability scores of the two ROIs.

\section{Results}

\subsection{Node optimization increases functional homogeneity of ROIs}
\label{sec:optimization_increases_homogeneity}

Construction of node-reconfiguring multilayer networks from fMRI data starts with optimizing ROIs used as network nodes. We do this separately for each subject and time window. Node optimization has two objectives: maximal functional homogeneity of ROIs and a narrow size distribution with only few extremely small or large ROIs. We define optimal ROIs using two variants of the greedy clustering algorithm introduced by Salehi et al. in~\cite{salehi2018exemplar} (dubbed weighted mean consistency and minimum correlation, according to the optimization target) and the Craddock algorithm \cite{craddock2012whole}, and compare them with ROIs from the Brainnetome atlas \cite{fan2016human} and random ROIs (for details on ROI definition, see section \ref{sec:node-definition}). All parcellation approaches yield some single-voxel ROIs, and the variants of the greedy clustering algorithm produce more of them than other approaches (Table~\ref{table:parcellation_props}). We consider these ROIs as noise or artifacts, and exclude them from all further analysis.

\begin{figure}
    \centering
    \includegraphics[width=0.9\textwidth]{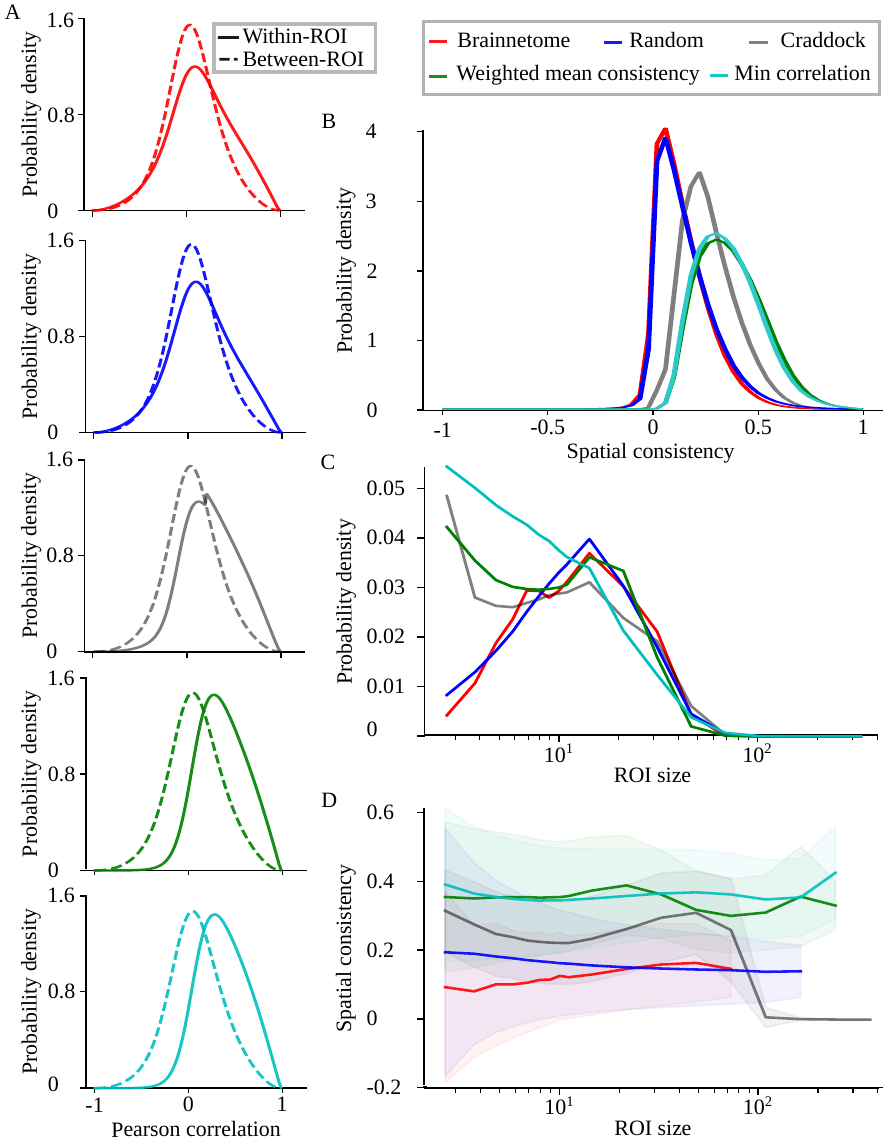}
    \caption{ROI optimization increases functional homogeneity. A) Distributions of Pearson correlation coefficient for voxels in the same ROI (Within-ROI) and in different ROIs (Between-ROI) for the five parcellation approaches considered. B) Distribution of spatial consistency. C) Distribution of ROI size (defined as the number of voxels in the ROI). D) Average spatial consistency (line) and its standard deviation (shaded area) as a function of ROI size.  All distributions have been calculated over all subjects, runs, time windows, and, in B, C, and D, ROIs, excluding ROIs that contain only a single voxel. In D, spatial consistencies have been binned based on ROI size and then bin-averaged.}
    \label{fig:corrs}
\end{figure}

To investigate ROIs' functional homogeneity, we calculate the Pearson correlation coefficient distributions between voxel pairs within the same ROI and in different ROIs (Figure \ref{fig:corrs}A). In a parcellation of homogeneous ROIs, the within-ROI correlations should be clearly stronger than the between-ROI ones. The average voxel-level correlation is stronger within the same ROI than between different ROIs for all parcellations tested (Table \ref{table:corrs-cons-values}). However, the difference of mean within-ROI and between-ROI correlations is small for Brainnetome and random ROIs, and the distributions of within-ROI and between-ROI correlations largely overlap for these parcellations. The right tail of the within-ROI distributions reaches further than that of the between-ROI distributions for all parcellations. However, the "excess" area between the within-ROI and between-ROI distributions where the within-ROI distribution is higher than the between-ROI distribution is substantially different for different parcellations (the larger the area, the more the within-ROI distribution is shifted towards high values relative to the between-ROI distribution). The excess area is more than twice as large for weighted mean consistency (0.386) and minimum correlation (0.392) approaches than for Brainnetome (0.155) or random (0.140) ROIs, while the excess area of the Craddock approach is located between these extremes (0.264) (Table \ref{table:in-roi_ref_excess_overlap}). The random ROIs do not, by definition, reflect any functional brain areas. Therefore, the similarity of Brainnetome and random ROIs suggests that Brainnetome ROIs may fail to match functional groups of voxels, and the increased within-ROI correlations may be due physical proximity of voxels within an ROI. 

Distributions of spatial consistency over subjects, runs, time windows, and ROIs (Figure \ref{fig:corrs}B) confirm that optimized ROIs are clearly more homogeneous than Brainnetome and random ROIs (Table \ref{table:corrs-cons-values}). The mean consistencies of weighted mean consistency and minimum correlation, 0.37 and 0.36, respectively, are more than twice that of Brainnetome's 0.14. Importantly, optimization increases not only the average homogeneity but also the number of ROIs with particularly high functional homogeneity, as demonstrated by the notable increase in the fraction of ROIs that have spatial consistency higher than 0.5 (Table \ref{table:corrs-cons-values}): 20 \% and 18 \% of ROIs for weighted mean consistency and minimum correlation, respectively, more than 10 times higher than the 1.4 \% for Brainnetome. This is possibly an even better measure for parcellation goodness than average consistency. This is because only substantially high consistency values indicate that ROIs match the underlying groups of correlated voxels: an increase in the consistency of less homogeneous ROIs from, for example, 0.1 to 0.2 increases average consistency but does not affect much the overall parcellation quality. Independently of ROI definition method, spatial consistency varies slightly over time (see Figure \ref{fig:SI_consistency_dynamics} in Supplementary Information). However, the changes are small and non-systematic, which justifies pooling time windows in Figure \ref{fig:corrs} and in the following analyses.

{}

\begin{table}[]
    \centering
    \begin{tabular}{l l l l l}
         Parcellation & N ROIs\textsuperscript{1} & N single-voxel ROIs\textsuperscript{1}  & size\textsuperscript{1, 2} & max ROI size\textsuperscript{1}  \\
         \hline
         Brainnetome & 245.95 $\pm$ 0.22 & 0.23 $\pm$ 0.54 & 20.70 $\pm$ 11.09 & 56.43 $\pm$ 5.74\\
         Filtered Brainnetome\textsuperscript{3} & 180.50 $\pm$ 9.19 & 0.22 $\pm$ 0.54 & 21.88 $\pm$ 10.98 & 56.09 $\pm$ 5.94\\
         Random & 246.00 $\pm$ 0.00 & 1.08 $\pm$ 1.06 & 20.76 $\pm$ 12.32 & 69.32 $\pm$ 11.69\\
         Craddock & 241.21 $\pm$ 4.34 & 0.59 $\pm$ 1.05 & 21.13 $\pm$ 15.31 & 100.39 $\pm$ 48.58\\
         Weighted mean consistency & 245.95 $\pm$ 0.22 & 23.21 $\pm$ 6.19 & 17.64 $\pm$ 10.77 & 60.69 $\pm$ 14.59 \\
         Min correlation & 245.95 $\pm$ 0.22 & 23.52 $\pm$ 6.23 & 17.57 $\pm$ 14.33 & 82.49 $\pm$ 18.99\\
         \hline
    \end{tabular}
    \caption{Basic properties of parcellations. \textsuperscript{1}: mean $\pm$ STD across subjects, runs, and layers, \textsuperscript{2}: without single-voxel ROIs, \textsuperscript{3}: for details, see section \ref{section:filtered-multiplex}. Note that the max ROI size given here does not match the largest ROI size value in Figs.~\ref{fig:corrs}C-D and \ref{fig:filtered_multiplex}B-C: the max size reported here is averaged across subjects, runs, and windows, while the largest ROI size value in those figures is the absolute maximum value present in the pooled dataset.}
    \label{table:parcellation_props}
\end{table}

\begin{table}[h!]
\centering
\begin{tabular}{ l l l l l l }
 & \multicolumn{2}{l}{Mean correlations} & \multicolumn{2}{l}{Spatial consistency} & \\
 & \multicolumn{2}{l}{of voxels excl. single-voxel ROIs} & \multicolumn{2}{l}{excl. single-voxel ROIs} & \\
 \cmidrule(lr){2-3} \cmidrule{4-6}
 & \multicolumn{1}{l}{} & \multicolumn{1}{l}{} & & & \multicolumn{1}{l}{\% of ROIs with} \\
Parcellation & \multicolumn{1}{l}{Within-ROI} & \multicolumn{1}{l}{Between-ROI} & \multicolumn{1}{l}{Mean\textsuperscript{1}} & \multicolumn{1}{l}{Median} & \multicolumn{1}{l}{consistency $> 0.5$} \\
\hline
Brainnetome & 0.15 & 0.059 & 0.14 $\pm$ 0.13 & 0.12 & 1.4 \\
Random & 0.15 & 0.059 & 0.16 $\pm$ 0.14 & 0.13 & 2.5 \\
Craddock & 0.26 & 0.059 & 0.27 $\pm$ 0.13 & 0.25 & 5.5 \\
Weighted mean consistency & 0.35 & 0.077 & 0.37 $\pm$ 0.15 & 0.35 & 20 \\
Min correlation & 0.36 & 0.077 & 0.36 $\pm$ 0.15 & 0.34 & 18 \\
\hline
\end{tabular}
\caption{Key indicators of correlations and consistencies in the parcellations. \textsuperscript{1}: mean $\pm$ STD across subjects, runs, layers, and ROIs.}
\label{table:corrs-cons-values}
\end{table}

\begin{table}[h!]
\centering
\begin{tabular}{c c c c c c}
\multirow{2}{*}{\includegraphics[width=0.1\textwidth]{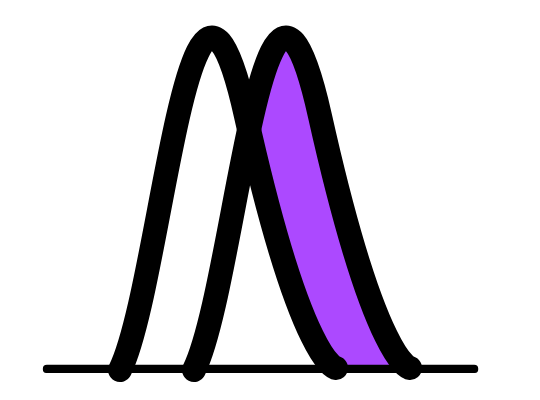}} & Brainnetome & Random & Craddock & Weighted mean consistency & Min correlation \\
\cline{2-6}
& & & & & \\
 & 0.155 & 0.140 & 0.264 & 0.386 & 0.392 \\
\hline
\end{tabular}
\caption{The area between the Within-ROI and Between-ROI curves of Figure \ref{fig:corrs}A for the part of the curves where Within-ROI is higher than Between-ROI (shaded area in illustration). A higher value indicates that the correlations are higher within ROIs than between ROIs.}
\label{table:in-roi_ref_excess_overlap}
\end{table}

Even after excluding the single-voxel ROIs, the greedy clustering algorithms (weighted mean consistency and minimum correlation) and the Craddock approach produce more small ROIs than the Brainnetome and random parcellations (Figure \ref{fig:corrs}C). This decreases the average ROI size for the greedy clustering algorithms (Table \ref{table:parcellation_props}). In the case of the Craddock approach, however, the increased number of small ROIs is compensated by the largest ROI that is larger than in any other parcellation (see Table \ref{table:parcellation_props}). This is because the Craddock algorithm produces new ROIs by cutting smaller components off from the largest connected component of the voxel-level network instead of dividing the largest component into two equally-sized parts. For ROI sizes larger than 10 voxels, the size distributions are similar for all parcellation approaches (see Figure~\ref{fig:corrs}C).

Importantly, the optimization approaches increase the spatial consistency of ROIs of all sizes (Figure \ref{fig:corrs}D). While the correlation between ROI size and consistency, calculated over subjects, runs, and layers, is significant due to the large number of data points, the correlation values are very small for all parcellation approaches excluding Brainnetome, where a weak positive correlation is present between spatial consistency and ROI size (Pearson $r$: 0.14, $p$: 0.0 for Brainnetome, $r$: -0.057, $p$: 0.0 for random ROIs, $r$: 0.078, $p$: 0.0 for Craddock, $r$: -0.016, $p$: $6.2 \times 10^{-138}$ for the weighted mean consistency approach, $r$: 0.018, $p$: $5.09 \times 10^{-191}$ for the minimum correlation approach). This indicates that the increased average spatial consistency of optimized ROIs is not due to an increased number of small ROIs that in some parcellations have higher spatial consistency than larger ROIs \cite{korhonen2017consistency}.

Subcortical ROIs of the Brainnetome parcellation are on average smaller than cortical ROIs and suffer from low signal-to-noise ratio (SNR). To ensure that the presence of these ROIs has not contributed to the lower mean spatial consistency of Brainnetome ROIs, we re-calculated the consistency and size distributions without subcortical ROIs. The results were similar to those obtained with the subcortical ROIs and ensured that the low spatial homogeneity of Brainnetome is not due to the weak SNR in the subcortical gray matter (Figure \ref{fig:SI_subcortical_areas}).

\subsection{Filtered multiplex networks suffer from nonhomogeneous nodes}
\label{section:filtered-multiplex}

The increased number of optimized ROIs with spatial consistency higher than 0.5 suggests that optimized ROIs match the underlying correlated voxel groups better than static ROIs. However, there is also another possible mechanism behind the increased functional homogeneity of optimized ROIs. Assuming that spatial consistency of (static) brain areas correlates with activity (see \cite{ryyppo2018regions}), ROI optimization would filter out inactive brain areas with low spatial consistency: due to thresholding, seed voxels located in inactive areas would never expand and would thus be removed from the parcellation as single-voxel ROIs. To assess the impact this mechanism has on the increase of functional homogeneity, we compared our node-reconfiguring multilayer network model to a simpler model, the filtered multiplex network. 

This model assumes that brain areas have static boundaries, but an area may be inactive and thus have no connections in certain windows. Therefore, all nodes are from a static parcellation but not all nodes are present on each layer. Unlike the node-reconfiguring model, the filtered multiplex model does not require re-optimizing ROIs for each time window and thus has lighter computational burden than the node-reconfiguring model.

We tested the filtered multiplex model using filtered versions of the Brainnetome parcellation that contained, in each time window, the $N$ ROIs with the highest spatial consistency. The lower $N$ is, the fewer ROIs there are on each layer, and consequently the fewer voxels from the set of all voxels are included. To assess whether filtering a static parcellation could yield a homogeneity increase similar to that caused by ROI optimization, we filtered ROIs so that the numbers of voxels matched between the filtered and optimized parcellations. That is, we selected the value of the free parameter $N$ so that the total number of voxels in the filtered ROIs was as high as possible without exceeding the number of voxels in the parcellation created by the weighted mean consistency approach. 

\begin{figure}
    \centering
    \includegraphics[width=\textwidth]{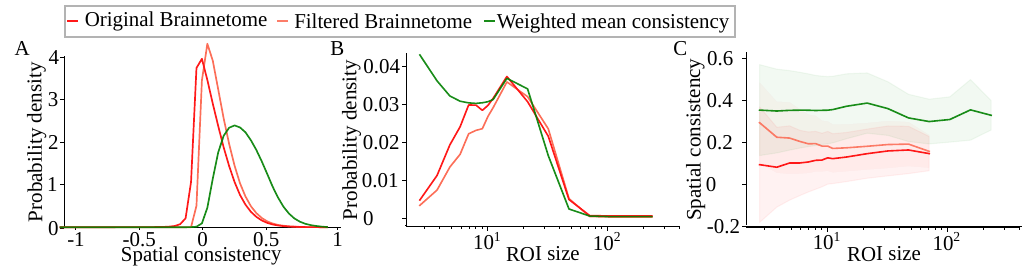}
    \caption{Average homogeneity of the filtered Brainnetome atlas is increased but does not reach the homogeneity of optimized ROIs. A) Distribution of spatial consistency in the original Brainnetome parcellation, the parcellation created with the weighted mean consistency approach, and the Brainnetome parcellation filtered based on the weighted mean consistency parcellation. B) Distribution of ROI size (defined as the number of voxels in the ROI). C) Average spatial consistency (line) and its standard deviation (shaded area) as a function of ROI size. All distribution have been calculated over all subjects, runs, time windows, and ROIs, excluding ROIs that contain only a single voxel. In C, spatial consistencies have been binned based on ROI size and then bin-averaged.}
    \label{fig:filtered_multiplex}
\end{figure}

Nodes of the filtered multiplex network have higher mean spatial consistency (0.19 $\pm$ 0.12), median of spatial consistency (0.16) and percentage of ROIs with spatial consistency larger than 0.5 (1.8 \%) than unfiltered Brainnetome ROIs (see Figure \ref{fig:filtered_multiplex}A). Meanwhile, filtering shifts the size distribution slightly to larger sizes (Figure \ref{fig:filtered_multiplex}B and Table \ref{table:parcellation_props}).

{}

Interestingly, filtering increases spatial consistency particularly in the smallest ROIs (Figure \ref{fig:filtered_multiplex}C). This, together with the weak but significant correlation between ROI size and spatial consistency, suggests that the least homogeneous Brainnetome ROIs are small. However, the large standard deviation of spatial consistency at smallest ROI sizes indicates that there are also small ROIs with high consistency. So, when ROIs with the lowest consistency are removed by filtering, spatial consistency of small ROIs is higher than that of larger ones. However, the correlation between ROI size and spatial consistency is very weak in the filtered data (Pearson $r$: 0.0077, $p$: $7.5 \times 10^{-29}$). Large consistency deviation and low consistency are not typical for small ROIs in the standard-space Brainnetome atlas \cite{korhonen2017consistency, ryyppo2018regions}. Therefore, we assume that the small ROIs with low consistency have suffered from a suboptimal registration to the native space, causing both decreased homogeneity and small size because of losing voxels.

Although filtering increases functional homogeneity, the ROIs of the filtered Brainnetome atlas are, on average, much less homogeneous than the ROIs optimized with the weighted mean consistency approach. This indicates that brain function requires reorganization of functional areas instead of activation and inactivation of static areas. Therefore, filtered multiplex networks cannot replace node-reconfiguring multilayer networks as a model of brain function.

\subsection{Node reoptimization between time windows is necessary for sustained high consistency}
\label{section:optimized-boundaries-change}

\begin{figure}
    \centering
    \includegraphics[width=0.75\textwidth]{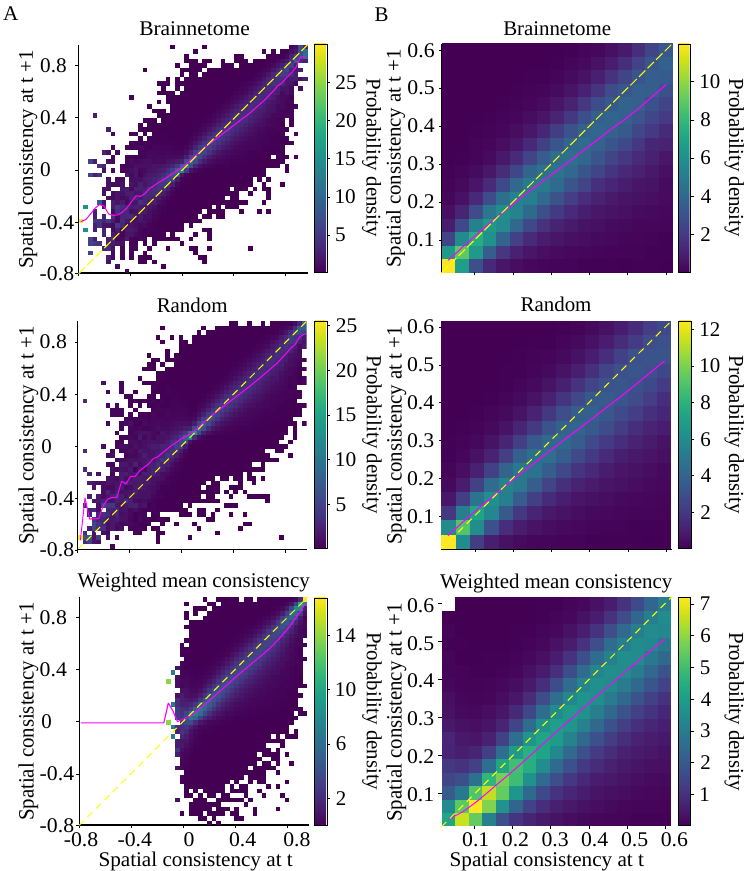}
    \caption{Optimal ROI boundaries are time-window-specific. Heatmaps of spatial consistency in time windows $t$ and $t + 1$ for Brainnetome atlas, random parcellation, and parcellation produced by the weighted mean consistency approach A) across the whole consistency range, and B) zoomed so that at least 90\% of data is included for all parcellations. Consistency is calculated with the same ROI boundaries in both windows (boundaries defined in window $t$ for random and weighted mean consistency parcellations, static ROIs for Brainnetome), and consistency distribution is calculated for each horizontal bin. Magenta line shows the average spatial consistency in each horizontal bin, and yellow dashed line corresponds to $x = y$. White cells contain no data. For corresponding heatmaps for Craddock and minimum correlation parcellations, see Figure \ref{fig:SI_consistency_in_next_window} in Supplementary Information.}
    \label{fig:consistency_in_next_window}
\end{figure}

The node-reconfiguring multilayer network leads to increased ROI homogeneity that the static Brainnetome ROIs or the filtered multiplex approach do not capture.
In the multilayer approach, the optimized ROIs change from one time window to the next.
To explore how crucial the reoptimization of ROIs from one window to the next is for consistency, we calculate spatial consistency in two consecutive time windows using the ROIs obtained \emph{in the first window} for both windows (see Figure \ref{fig:consistency_in_next_window}). That is, we take ROIs defined in window $t$ and use those ROI boundaries on data in window $t+1$ in addition to window $t$.
Note that, in the case of the Brainnetome atlas, where the ROIs are static, the same ROIs are always used for both windows. The drastic difference between data-driven and static ROIs is clearly visible in Figure \ref{fig:consistency_in_next_window}: while Brainnetome and random parcellations contain several ROIs with spatial consistency lower than zero, no such ROIs are present in the optimized parcellations.

In all optimized parcellations, ROIs have, on average, notably higher consistency in the window where they have been defined than in the following window (Table \ref{table:consistency_in_next_window}). If this was caused by random fluctuations in voxel time series similarity, similar change should be visible also in Brainnetome and random ROIs. However, this is not the case: the mean relative change in consistency for Brainnetome and random ROIs is close to zero. This indicates that functionally homogeneous brain areas reconfigure systematically also at short time scales of time windows, not only between cognitive tasks or different subjects. This means that ROI optimization is window-specific, and has to be done for each window separately.

\begin{table}[]
    \centering
    \begin{tabular}{l l l}
         Parcellation & Mean absolute change & Mean relative change\textsuperscript{1}\\
         \hline
         Brainnetome & -4.24$\times10^{-6} \pm$ 0.075 & -2.97$\times10^{-5}$\\
         Random & 2.56$\times10^{-5} \pm$ 0.083 & 1.63$\times10^{-4}$\\
         Craddock & -0.047 $\pm$ 0.10 & -0.18\\
         Weighted mean consistency & -0.059 $\pm$ 0.11 & -0.16\\
         Min correlation & -0.057 $\pm$ 0.12 & -0.16\\
         \hline
    \end{tabular}
    \caption{Relative change in spatial consistency between time windows $t$ and $t+1$, when ROIs defined in window $t$ are used for both windows. Negative change indicates higher consistency in window $t$. All averages have been calculated excluding the single-voxel ROIs. \textsuperscript{1}: mean absolute change divided by mean spatial consistency in window $t$.}
    \label{table:consistency_in_next_window}
\end{table}

{}

Interestingly, the spatial consistency of optimized ROIs decreases between consecutive time windows independently of the consistency value in the first window (Figure \ref{fig:consistency_in_next_window}). On the other hand, spatial consistency of Brainnetome and random ROIs is lower in the second window only if the consistency in the first window is close to the maximum consistency value. This can be explained by purely statistical reasons. Most ROIs have intermediate consistency, and high consistency values are rare (Figure \ref{fig:corrs}B). Let us assume that the consistency of each ROI in the second window is assigned randomly; this is a reasonable assumption at least for the random parcellation. If the consistency of ROI $I$ in the first window, $\phi_I^t$, is high, there are only few higher consistency values available in the second window, and therefore most probably $\phi_I^{t+1} < \phi_I^t$. Similarly, if $\phi_I^t$ is particularly low, probably $\phi_I^{t+1} > \phi_I^{t}$, which is also true for the Brainnetome and random ROIs.

\subsection{ROIs reorganize in a non-trivial way}
\label{section:ROI-reorganization}

\begin{figure}
    \centering
    \includegraphics[width=0.49\textwidth]{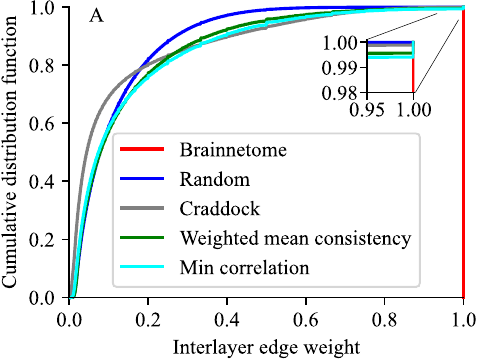}
\hfill
    \includegraphics[width=0.49\textwidth]{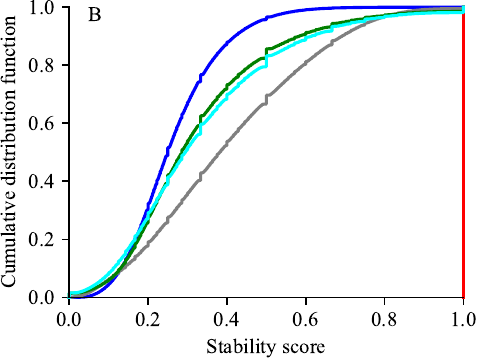}
\caption{ROI reconfiguration behavior is complex. Cumulative distribution functions of A) interlayer edge weights and B) stability scores, calculated over subjects, runs, and time windows, for different ROI definition methods, excluding ROIs that contain only a single voxel.}
\label{fig:interlayer_weights_and_stabilities}
\end{figure}

To investigate the reorganization of ROI boundaries, we construct the distributions of interlayer edge weights, that is, the Jaccard indices of ROIs from one time window to the next, for different ROI definition methods (Figure \ref{fig:interlayer_weights_and_stabilities}). Most weights are small for all ROI definition methods except Brainnetome. This is to be expected: the light-weight interlayer edges reflect small fluctuations of a few voxels in ROI boundaries, and these small fluctuations are more common than larger reorganizations of more stable voxel clusters that yield heavier edges.
Brainnetome ROIs do not change over time, and consequently all interlayer edge weights are equal to 1. Random ROIs have the lightest tail of the distribution, meaning that the random assignment produces the least amount of highly stable ROIs from one layer to the next, as expected when there is no regard to underlying data in the ROI definition. Craddock has the quickest growing early curve, meaning it has the highest proportion of small weights, even more than random ROI assignment. The distributions for the weighted mean consistency and minimum correlation approaches are close to each other, and both have a small jump at weight 1, meaning that there are a little below 1 \% of ROIs that stay exactly the same from one layer to the next. Craddock, weighted mean consistency, and minimum correlation also have a higher proportion of edge weights greater than approximately 0.15 than random ROI assignment, meaning that they exhibit a degree of ROI stability not explained by random behavior. The reorganization behavior of ROIs is complex: there is variation and reconfiguration of various magnitudes, and at the same time in the greedy growth data-driven ROI definition methods (weighted mean consistency and minimum correlation) there are even some fully stable ROIs.

The stability score of an ROI is the weight of the highest-weight interlayer edge connecting the ROI to another ROI on the next layer, and it reflects how stable the ROI stays in the layer transition (Figure \ref{fig:interlayer_weights_and_stabilities}, for the definition of the stability score see section \ref{section:network_analysis}).
Again, all Brainnetome ROIs are fully stable by definition, and random ROI assignment has only few highly stable ROIs. Out of the node-reconfiguring methods, Craddock ROIs have the highest median stability score; however, the weighted mean consistency and minimum correlation approaches yield the highest amount of highly stable ROIs (they have more ROIs with stability score $\gtrapprox 0.75$, the crossover point of the curves). They also have fully unstable ROIs (stability score 0), a result of voxel thresholding: some voxels do not fit into any ROIs and are left out, and consequently some ROIs completely disappear.
The most stable ROIs have variation in their sizes: among ROIs with stability score 1, there are many different ROI sizes present (see Table \ref{table:max_stability_roi_size_distribution} in Supplementary Information).
Therefore, the ROI landscape encompasses the full spectrum from fully unstable to fully stable ROIs, with the majority having various levels of stability. ROI reorganization is therefore not a trivial process, and defining ROIs as homogeneous units within the brain has to take this temporal reorganization into account.

\begin{figure}
    \centering
    \includegraphics[width=\textwidth]{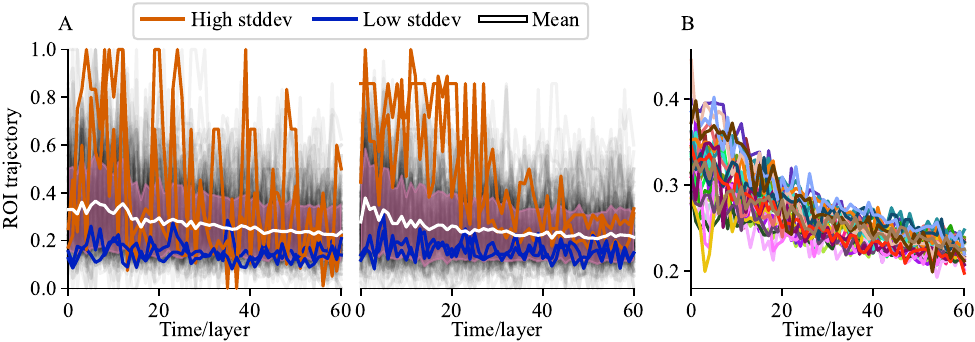}
    \caption{ROIs can change and return to original configuration on time scales longer than time windows, as evidenced by their trajectories. The trajectory of an ROI is the largest Jaccard index of that ROI with another ROI out of all ROIs on a temporally subsequent layer, which forms a time series of ROI stability and fluctuation. A high trajectory means the ROI stays stable; a low trajectory means the ROI has disappeared (voxels dispersed into other ROIs). A) The trajectories of all ROIs with at least 5 voxels from layer 0 for two representative subjects (left and right panels) for an imaging run with the weighted mean consistency method. All trajectories are shown in light gray, the mean trajectory is shown in white with pink standard deviation bands around, the two trajectories with highest standard deviation are shown in orange, and the two trajectories with the lowest standard deviation are shown in blue. B) The mean trajectories for all subjects for the imaging run.}
    \label{fig:trajectories_run4}
\end{figure}

Optimized ROIs change (and return to original configurations) also on time scales longer than time windows, still within the same cognitive task.
The trajectories of the ROIs reflect this change (for definition of trajectories, see section \ref{section:network_analysis}). Figure \ref{fig:trajectories_run4} shows all ROI trajectories for two subjects and mean trajectories for all subjects for one imaging run. In the high-variance trajectories, we see that for one subject one of the ROIs disappears but then returns exactly as it was around layer 39. On the other hand, for the other subject, the ROIs stay moderately stable until around layer 30, and then the ROI ensemble reorganizes and the previously stable high-variance ROIs do not appear again. The mean trajectories also differ from subject to subject, but all quickly decrease with the increasing time difference to the reference layer. As can be seen from the first example subject, there are however also ROIs that are highly similar in the first window and in later, even the last window. Such ROIs are not exceptionally stable but return close to their original configuration after significant reorganization. This suggests that the observed reorganization of ROIs is not random or caused by clustering noise but is a genuine feature of brain function, possibly reflecting changes in the naturalistic stimulus and subjects' reactions to it. The example illustrates that there is individual variability in ROI stability, and even high differences in ROI stability within individuals.

\subsection{Maximum values of intralayer connectivity are reached with non-extreme values of interlayer stability}
\label{section:intra-and-interlayer-properties}

\begin{figure}
    \centering
    \includegraphics[width=0.98\textwidth]{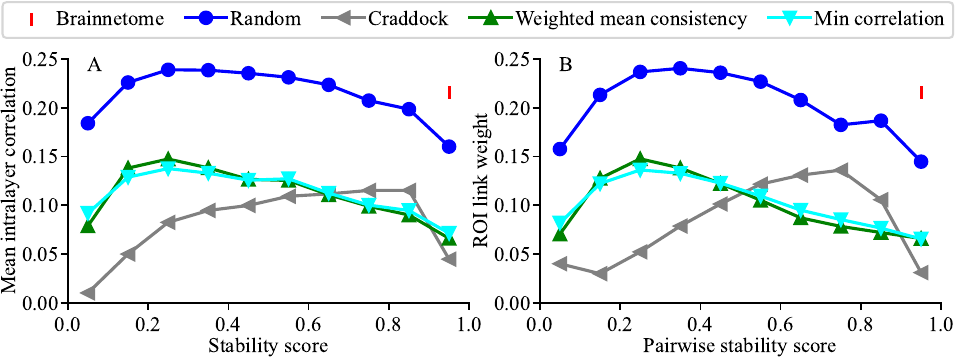}
\caption{In node-reconfiguring networks, maximal mean intralayer correlations and maximal link weights between ROIs occur at ROIs that are neither fully stable nor fully unstable. A) Mean correlation of an ROI to other ROIs within a layer versus ROI stability score. B) Link weight (i.e. correlation) between two ROIs within a layer versus pairwise stability score, calculated as the mean of the stability scores of the two ROIs. In both A) and B), the points are mean values averaged over all subjects, runs, and layers, binned in bins of size 0.1 (first bin $[0,0.1]$, second bin $(0.1,0.2]$, and so on), excluding ROIs that contain only a single voxel.}
\label{fig:stability_corr_link_weight}
\end{figure}

An ROI on a layer is connected to other ROIs on the same layer (intralayer edges, functional connectivity) and to ROIs on the subsequent layer (interlayer edges, ROI reorganization). Next, we investigate the interdependence of these two  connectivity types. The intralayer network is a complete network, where each ROI is connected to all other ROIs on that layer via edges defined in terms of correlation between ROI time series. A measure of an ROI's intralayer connectivity is the mean correlation of that ROI to the other ROIs (i.e. mean intralayer edge weight), while the stability score measures ROI's interlayer behavior. For all ROI optimization methods as well as for random ROIs, the mean intralayer correlation versus stability score forms an approximately inverted-"U"-shaped curve, while the static Brainnetome ROIs always have a stability score of 1 and appear only as a single data point instead of a curve (Figure \ref{fig:stability_corr_link_weight}A). For the greedy growth ROI definition methods (weighted mean consistency, minimum correlation, and random), the highest mean intralayer correlation co-occurs with stability scores between 0.2 and 0.3, while for the min-cut-based Craddock method the highest mean intralayer correlation co-occurs with stability scores between 0.8 and 0.9.

There is a clear division of the methods into two classes based on their intralayer correlation level: random and Brainnetome have intralayer correlations larger than Craddock, weighted mean consistency, and minimum correlation. This behavior reflects the higher functional homogeneity of the ROIs in the latter class of methods: when ROIs are more homogeneous, their "leftover" correlations to the other ROIs are smaller than if they were less homogeneous.

Another measure of intralayer ROI similarity is the intralayer edge weight, that is, the magnitude of the correlation between two ROIs. Figure \ref{fig:stability_corr_link_weight}B shows the edge weight (link weight) versus the pairwise ROI stability score for the different ROI definition methods. Again, random and Brainnetome ROIs have higher edge weights than the other methods. The behavior of Craddock versus weighted mean consistency and minimum correlation is nearly mirrored in the horizontal direction: the \emph{minimum} edge weight for Craddock lies a little below stability 0.2, and the \emph{maximum} edge weight for weighted mean consistency and minimum correlation lies a little above 0.2. In contrast, the \emph{maximum} edge weight for Craddock lies a little below 0.8, and the \emph{minimum} edge weight for weighted mean consistency and minimum correlation is reached at the maximum stability around 1.0.

The highest mean correlations and edge weights are achieved with ROIs that are neither fully unstable nor fully stable. That is, there are ROIs of intermediate stability that form the strongest connections to other ROIs. 
Interestingly, there is a marked difference between Craddock (min-cut) and greedy growth ROI definition methods in how stable the strongest-connected ROIs are.

\section{Discussion}

Our proposed node-reconfiguring network model for human brain function yields significant increases in the functional homogeneity of Regions of Interest (ROIs). Our model is able to capture short-scale changes in the boundaries of homogeneous voxel groups, and our results show that these boundaries indeed change in time, also at short time scales of $\sim$10 s, notably shorter than the timescales investigated in earlier studies \cite{iraji2019spatialdynamics, salehi2020there, luo2021within}. Until recently, the network neuroscience community has tried to mitigate the problem of low functional homogeneity by introducing more elaborated static parcellation approaches. However, our results show that this approach will not solve the problem: \textit{any} static parcellation approach unavoidably leads to low functional homogeneity and data losses and thus cannot produce an accurate network model of brain function.

\subsection{Functionally homogeneous brain areas reorganize at short time scales}

Our results on the low functional homogeneity of static ROI atlases traditionally used as functional brain network nodes are consistent with previous literature \cite{korhonen2017consistency, ryyppo2018regions, gottlich2013altered, stanley2013defining}. However, different ROI optimization approaches, for example the Craddock \cite{craddock2012whole} and greedy clustering \cite{salehi2018exemplar} algorithms, can significantly increase the functional homogeneity of ROIs. 

There are two possible explanations for the low functional homogeneity of ROIs in static atlases. First, if the underlying brain function is based on a static set of functionally homogeneous areas, ROI boundaries may not match these areas. In this case, a parcellation approach may improve functional homogeneity by introducing a new set of static ROIs that better matches the underlying areas. Until recently, most attempts to mitigate the problem of low functional homogeneity have aimed at this. However, our results support another explanation for low functional homogeneity: the boundaries of the underlying homogeneous areas change over time, which makes it impossible to detect optimally homogeneous ROIs with any static parcellation. We have shown that filtering the static Brainnetome atlas by removing the least homogeneous ROIs does not increase functional homogeneity similarly as using ROIs with time-dependent boundaries. This indicates that reorganization of the underlying brain areas happens through changes in the boundaries of all ROIs instead of activation and inactivation of static ROIs. 

Earlier literature has reported reorganization of ROI boundaries between subjects and within a subject between cognitive tasks \cite{luo2022inside, salehi2020there, salehi2020individualized, kong2019spatial}.
Our results show that brain function requires reorganization of ROI boundaries at multiple time scales, ranging from fast changes from one time window (duration 8 seconds) to the next to slower fluctuations over several time windows. These fluctuations in ROI boundaries may reflect changes in the naturalistic free story-listening task.

\subsection{Multilayer or multiplex brain networks?}

We propose the node-reconfiguring multilayer network model that combines consecutive time windows into a single functional brain network using the optimized ROIs as nodes on each layer. The important difference between our model and the existing time-dependent multilayer network models of brain function (e.g. \cite{bassett2011dynamic, bassett2013task, telesford2017cohesive, braun2015dynamic}) is the definition of nodes and interlayer edges: earlier models have constructed multiplex networks where nodes are static ROIs and only diagonal interlayer edges with unit weight are allowed. Our model, to the contrary, uses time-dependent ROIs as nodes and allows interlayer edges between all nodes. The weight of these edges varies, depicting the fraction of shared voxels between two ROIs. While the multiplex models can capture the reorganization in networks of ROIs \cite{bassett2011dynamic}, they omit the reorganization at the more fine spatial scale of ROI boundaries.  

Buld\'u \& Porter \cite{buldu2018frequency} compared multiplex and multilayer models for connections within and across MEG frequency bands. They showed that theoretically predicted algebraic connectivity of multilayer networks matches the data better than that of multiplex networks. Further, they, in general, recommended multilayer models instead of multiplex ones because the connections between frequency bands are never guaranteed to be limited to the diagonal ones. This criticism applies also to networks where layers correspond to time windows \cite{buldu2018frequency}. 

To our knowledge, our model is the first multilayer brain network model with time-dependent nodes. In single-layer network analysis, simulations \cite{yu2017comparing} have shown that static nodes may cause spuriosities in the observed network structure compared to time-dependent nodes. The first application of data-driven node definition approaches on real neuroimaging data \cite{luo2021within} confirmed that time-dependent nodes open a richer view of brain connectivity than static ones. The node definition approach of \cite{luo2021within} is built on the same greedy clustering algorithm \cite{salehi2018exemplar} as our weighted mean consistency and minimum correlation approaches. However, the nodes of \cite{luo2021within} varied at the slower time scale, that is, between cognitive tasks.

Many analysis methods tailored for multilayer networks require or at least benefit from the existence of non-diagonal interlayer links. For example, examination of the interplay of intra- and interlayer network properties is possible only when the networks have non-trivial intra- and interlayer structure. We have performed a rudimentary example analysis in section \ref{section:intra-and-interlayer-properties}. We find that the most-connected ROIs, as well as ROI pairs sharing the heaviest intralayer edges, are neither fully stable nor fully unstable, a fact not possible to uncover with static-parcellation multiplex networks. Additionally, the trajectories, or temporal reorganization behaviors, of ROIs are also impossible to investigate with static-parcellation multiplex models of the brain (see section \ref{section:ROI-reorganization}). If the brain behaves as a multiplex network, the multilayer network formalism allows multiplex behavior to emerge (ROIs can stay stable); however, when the brain does not behave as a multiplex network, a multiplex network model cannot capture the richness of phenomena and analysis allowed by the multilayer formulation.

\subsection{Multilayer networks as models of spatiotemporal brain dynamics}

In addition to increasing the accuracy of functional network analysis through improved functional homogeneity of ROIs and the rich landscape of interlayer edges, our node-reconfiguring multilayer network model contributes to a more fundamental paradigm shift in network neuroscience, the rise of the chronnectome paradigm. Traditionally, the neuroscience community saw functional connectivity as a set of static similarities between static brain areas. Nowadays, we know that functional edge weights change at several time scales, including at short scales between cognitive tasks or spontaneously in rest (for review, see \cite{calhoun2014chronnectome, korhonen2021principles}), and also fluctuations in network properties have been reported since early simulation studies (e.g. \cite{honey2007network}). However, the stasis of network nodes seems to remain as a cornerstone of our understanding of functional connectivity. The chronnectome paradigm \cite{calhoun2014chronnectome, iraji2019spatialchronnectome, iraji2020space} questions this view and describes the brain as a collection of sources with continuously reconfiguring spatial and connectivity profiles.

Adopting the chronnectome paradigm requires not only a change in the way of thinking about functional connectivity, but also significant development of new analysis tools to explore sources that change in terms of space, time, and connectivity \cite{iraji2020space}. This is a huge requirement, and therefore it is not surprising that many chronnectome studies have so far concentrated on localizing the sources in space and time (see e.g. \cite{wu2018approach, wu2021tracking, iraji2019spatialdynamics, iraji2019spatialchronnectome, salehi2020individualized, salehi2020there}), while network analyses that would use sources as nodes are rare (\cite{luo2021within} is the only one we are aware of). Our node-reconfiguring multilayer network model offers an excellent framework for exploring the spatiotemporal dynamics of the brain.

\subsection{Future work}

Building the multilayer networks with optimized nodes requires several methodological choices, including the selection of node optimization and edge definition methods. In the present article, we have used the greedy clustering and Craddock algorithms to obtain time-dependent nodes and have defined interlayer edges in terms of ROI overlap. Intralayer edge weights correspond to the Pearson correlation coefficient, which, despite its known problems, is the most commonly used similarity measure in network studies of fMRI data \cite{smith2011network, zalesky2012use, mehler2018lure}. However, our approach to constructing multilayer networks does not depend on these methodological choices. Therefore, the modular nature of the model and its Python implementation allow any node or edge definition to be embedded in the model without changing the overall framework. In particular, the accuracy of functional network analysis would probably benefit from exploring more advanced edge definition methods, for example the multivariate distance correlation \cite{geerligs2016functional}.

The remarkable increase in functional homogeneity due to window-wise ROI optimization possibly comes with the cost of unrealistically low temporal stability of ROIs. While earlier results on voxel-level connectivity changes at short time scales \cite{ryyppo2018regions, luo2022inside} suggest that ROI boundaries should fluctuate to some extent, successful brain function probably requires a compromise between complete stability and excessive short-scale changes. In addition, extreme fluctuations in ROI boundaries make group-level analysis and setting results in anatomical context more difficult. As a possible solution, one could apply clustering methods such as multislice modularity maximization \cite{mucha2010community} or multilayer stochastic block models \cite{pamfil2019relating} on voxel-level temporal (multiplex) networks where interlayer links are allowed only between replicas of the same voxel on consecutive layers. These methods could produce relatively stable ROIs without compromising functional homogeneity or the ability to detect functionally meaningful fluctuations.

Modeling the brain as a multilayer network with time-dependent nodes opens possibilities to apply previously unaccessible network analysis methods. In section \ref{section:intra-and-interlayer-properties}, we performed an example analysis of the interplay between the properties of intra- and interlayer networks. Further analysis methods tailored for multilayer networks include multilayer motif analysis \cite{battiston2017multilayer} and multilayer clustering approaches \cite{mucha2010community, stanley2016clustering}. Multilayer motif analysis detects patterns that appear in the network more commonly than expected if the network were randomly wired. Multilayer clustering approaches detect groups of nodes that are tightly connected both within time windows and over the time dimension.

In addition to multilayer network analysis, optimizing ROIs separately for each time window opens another possible research direction: clustering time windows into temporal states based on their ROI configuration. The hypothesis of brain function as a collection of metastable states and abrupt transitions between them was originally proposed in the EEG microstate literature \cite{custo2017electroencephalographic, khanna2015microstates}, and previous fMRI studies using, e.g., independent component analysis (ICA) or co-activation patterns (CAP) methodology have supported this hypothesis \cite{iraji2019spatialchronnectome, iraji2019spatialdynamics, long2021graph, liu2018co, hutchison2013dynamic}. Furthermore, the outcomes of our trajectory analysis, especially the observation of ROIs that return to their original configuration after several windows, suggest that reconfiguration of functional connectivity takes place as transitions between a limited set of states rather than as random fluctuations.

\section*{Code availability}

The node-reconfiguring multilayer network model, the node optimization approaches, and other code used for the analysis are available as a Python implementation at \url{https://github.com/ercco/multilayer-brains}.

\section*{Acknowledgements}

We thank Prof. Iiro J\"a\"askel\"ainen for sharing the dataset used in this study and for assistance in data-related issues and Jari Saram\"aki for constructive discussions on the project. PDL acknowledges funding from the Erasmus+ program. MH acknowledges funding from the National Institute On Deafness And Other Communication Disorders of the National Institutes of Health (NIDCD/NIH) (K99DC022305). OK acknowledges funding from Emil Aaltonen Foundation (190095), Osk. Huttunen Foundation, and Academy of Finland (current Research Council of Finland; 332368). TN and MK acknowledge funding from the Research Council of Finland (353799 and 349366).

\bibliographystyle{plain}
\bibliography{references}

\begin{thebibliography}{10}

\bibitem{alakorkko2017effects}
Tuomas Alak{\"o}rkk{\"o}, Heini Saarim{\"a}ki, Enrico Glerean, Jari Saram{\"a}ki, and Onerva Korhonen.
\newblock Effects of spatial smoothing on functional brain networks.
\newblock {\em European Journal of Neuroscience}, 46(9):2471--2480, 2017.

\bibitem{andellini2015test}
Martina Andellini, Vittorio Cannat{\`a}, Simone Gazzellini, Bruno Bernardi, and Antonio Napolitano.
\newblock Test-retest reliability of graph metrics of resting state mri functional brain networks: A review.
\newblock {\em Journal of Neuroscience Methods}, 253:183--192, 2015.

\bibitem{aurich2015evaluating}
Nathassia~K Aurich, Jos{\'e}~O Alves~Filho, Ana~M Marques~da Silva, and Alexandre~R Franco.
\newblock Evaluating the reliability of different preprocessing steps to estimate graph theoretical measures in resting state {fMRI} data.
\newblock {\em Frontiers in Neuroscience}, 9:48, 2015.

\bibitem{bassett2017network}
Danielle~S Bassett and Olaf Sporns.
\newblock Network neuroscience.
\newblock {\em Nature Neuroscience}, 20(3):353--364, 2017.

\bibitem{bassett2011dynamic}
Danielle~S Bassett, Nicholas~F Wymbs, Mason~A Porter, Peter~J Mucha, Jean~M Carlson, and Scott~T Grafton.
\newblock Dynamic reconfiguration of human brain networks during learning.
\newblock {\em Proceedings of the National Academy of Sciences}, 108(18):7641--7646, 2011.

\bibitem{bassett2013task}
Danielle~S Bassett, Nicholas~F Wymbs, M~Puck Rombach, Mason~A Porter, Peter~J Mucha, and Scott~T Grafton.
\newblock Task-based core-periphery organization of human brain dynamics.
\newblock {\em PLoS Computational Biology}, 9(9):e1003171, 2013.

\bibitem{battiston2017multilayer}
Federico Battiston, Vincenzo Nicosia, Mario Chavez, and Vito Latora.
\newblock Multilayer motif analysis of brain networks.
\newblock {\em Chaos: An Interdisciplinary Journal of Nonlinear Science}, 27(4):047404, 2017.

\bibitem{boccaletti2014structure}
Stefano Boccaletti, Ginestra Bianconi, Regino Criado, Charo~I Del~Genio, Jes{\'u}s G{\'o}mez-Gardenes, Miguel Romance, Irene Sendina-Nadal, Zhen Wang, and Massimiliano Zanin.
\newblock The structure and dynamics of multilayer networks.
\newblock {\em Physics Reports}, 544(1):1--122, 2014.

\bibitem{braun2015dynamic}
Urs Braun, Axel Sch{\"a}fer, Henrik Walter, Susanne Erk, Nina Romanczuk-Seiferth, Leila Haddad, Janina~I Schweiger, Oliver Grimm, Andreas Heinz, Heike Tost, et~al.
\newblock Dynamic reconfiguration of frontal brain networks during executive cognition in humans.
\newblock {\em Proceedings of the National Academy of Sciences}, 112(37):11678--11683, 2015.

\bibitem{brookes2016multi}
Matthew~J Brookes, Prejaas~K Tewarie, Benjamin~AE Hunt, Sian~E Robson, Lauren~E Gascoyne, Elizabeth~B Liddle, Peter~F Liddle, and Peter~G Morris.
\newblock A multi-layer network approach to {MEG} connectivity analysis.
\newblock {\em NeuroImage}, 132:425--438, 2016.

\bibitem{buldu2018frequency}
Javier~M Buld{\'u} and Mason~A Porter.
\newblock Frequency-based brain networks: From a multiplex framework to a full multilayer description.
\newblock {\em Network Neuroscience}, 2(4):418--441, 2018.

\bibitem{calhoun2014chronnectome}
Vince~D Calhoun, Robyn Miller, Godfrey Pearlson, and Tulay Adal{\i}.
\newblock The chronnectome: time-varying connectivity networks as the next frontier in {fMRI} data discovery.
\newblock {\em Neuron}, 84(2):262--274, 2014.

\bibitem{craddock2012whole}
R~Cameron Craddock, G~Andrew James, Paul~E Holtzheimer~III, Xiaoping~P Hu, and Helen~S Mayberg.
\newblock A whole brain {fMRI} atlas generated via spatially constrained spectral clustering.
\newblock {\em Human Brain Mapping}, 33(8):1914--1928, 2012.

\bibitem{crofts2016structure}
Jonathan~J Crofts, Michael Forrester, and Reuben~D O'Dea.
\newblock Structure-function clustering in multiplex brain networks.
\newblock {\em EPL (Europhysics Letters)}, 116(1):18003, 2016.

\bibitem{ccukur2013attention}
Tolga {\c{C}}ukur, Shinji Nishimoto, Alexander~G Huth, and Jack~L Gallant.
\newblock Attention during natural vision warps semantic representation across the human brain.
\newblock {\em Nature Neuroscience}, 16(6):763--770, 2013.

\bibitem{custo2017electroencephalographic}
Anna Custo, Dimitri Van De~Ville, William~M Wells, Miralena~I Tomescu, Denis Brunet, and Christoph~M Michel.
\newblock Electroencephalographic resting-state networks: source localization of microstates.
\newblock {\em Brain Connectivity}, 7(10):671--682, 2017.

\bibitem{de2017multilayer}
Manlio De~Domenico.
\newblock Multilayer modeling and analysis of human brain networks.
\newblock {\em Giga Science}, 6(5):gix004, 2017.

\bibitem{eickhoff2015connectivity}
Simon~B Eickhoff, Bertrand Thirion, Ga{\"e}l Varoquaux, and Danilo Bzdok.
\newblock Connectivity-based parcellation: Critique and implications.
\newblock {\em Human Brain Mapping}, 36(12):4771--4792, 2015.

\bibitem{fan2016human}
Lingzhong Fan, Hai Li, Junjie Zhuo, Yu~Zhang, Jiaojian Wang, Liangfu Chen, Zhengyi Yang, Congying Chu, Sangma Xie, Angela~R Laird, et~al.
\newblock The human {B}rainnetome atlas: a new brain atlas based on connectional architecture.
\newblock {\em Cerebral cortex}, 26(8):3508--3526, 2016.

\bibitem{feinberg2013ultra}
David~A Feinberg and Kawin Setsompop.
\newblock Ultra-fast {MRI} of the human brain with simultaneous multi-slice imaging.
\newblock {\em Journal of Magnetic Resonance}, 229:90--100, 2013.

\bibitem{fornito2013graph}
Alex Fornito, Andrew Zalesky, and Michael Breakspear.
\newblock Graph analysis of the human connectome: promise, progress, and pitfalls.
\newblock {\em NeuroImage}, 80:426--444, 2013.

\bibitem{geerligs2016functional}
Linda Geerligs, Cam-CAN, and Richard~N Henson.
\newblock Functional connectivity and structural covariance between regions of interest can be measured more accurately using multivariate distance correlation.
\newblock {\em NeuroImage}, 135:16--31, 2016.

\bibitem{gottlich2013altered}
Martin G{\"o}ttlich, Thomas~F M{\"u}nte, Marcus Heldmann, Meike Kasten, Johann Hagenah, and Ulrike~M Kr{\"a}mer.
\newblock Altered resting state brain networks in {P}arkinson’s disease.
\newblock {\em PloS one}, 8(10):e77336, 2013.

\bibitem{guillon2017loss}
Jeremy Guillon, Yohan Attal, Olivier Colliot, Valentina La~Corte, Bruno Dubois, Denis Schwartz, Mario Chavez, and F~De~Vico Fallani.
\newblock Loss of brain inter-frequency hubs in {A}lzheimer's disease.
\newblock {\em Scientific Reports}, 7(1):10879, 2017.

\bibitem{hakonen2022processing}
Maria Hakonen, Arsi Ik{\"a}heimonen, Annika Hult{\`e}n, Janne Kauttonen, Miika Koskinen, Fa-Hsuan Lin, Anastasia Lowe, Mikko Sams, and Iiro~P J{\"a}{\"a}skel{\"a}inen.
\newblock Processing of an audiobook in the human brain is shaped by cultural family background.
\newblock {\em Brain Sciences}, 12(5):649, 2022.

\bibitem{honey2007network}
Christopher~J Honey, Rolf K{\"o}tter, Michael Breakspear, and Olaf Sporns.
\newblock Network structure of cerebral cortex shapes functional connectivity on multiple time scales.
\newblock {\em Proceedings of the National Academy of Sciences}, 104(24):10240--10245, 2007.

\bibitem{hsu2017simultaneous}
Yi-Cheng Hsu, Ying-Hua Chu, Shang-Yueh Tsai, Wen-Jui Kuo, Chun-Yuan Chang, and Fa-Hsuan Lin.
\newblock Simultaneous multi-slice inverse imaging of the human brain.
\newblock {\em Scientific Reports}, 7(1):17019, 2017.

\bibitem{hutchison2013dynamic}
R~Matthew Hutchison, Thilo Womelsdorf, Elena~A Allen, Peter~A Bandettini, Vince~D Calhoun, Maurizio Corbetta, Stefania Della~Penna, Jeff~H Duyn, Gary~H Glover, Javier Gonzalez-Castillo, et~al.
\newblock Dynamic functional connectivity: promise, issues, and interpretations.
\newblock {\em NeuroImage}, 80:360--378, 2013.

\bibitem{iraji2019spatialchronnectome}
Armin Iraji, Thomas~P Deramus, Noah Lewis, Maziar Yaesoubi, Julia~M Stephen, Erik Erhardt, Aysneil Belger, Judith~M Ford, Sarah McEwen, Daniel~H Mathalon, et~al.
\newblock The spatial chronnectome reveals a dynamic interplay between functional segregation and integration.
\newblock {\em Human Brain Mapping}, 40(10):3058--3077, 2019.

\bibitem{iraji2019spatialdynamics}
Armin Iraji, Zening Fu, Eswar Damaraju, Thomas~P DeRamus, Noah Lewis, Juan~R Bustillo, Rhoshel~K Lenroot, Aysneil Belger, Judith~M Ford, Sarah McEwen, et~al.
\newblock Spatial dynamics within and between brain functional domains: A hierarchical approach to study time-varying brain function.
\newblock {\em Human Brain Mapping}, 40(6):1969--1986, 2019.

\bibitem{iraji2020space}
Armin Iraji, Robyn Miller, Tulay Adali, and Vince~D Calhoun.
\newblock Space: a missing piece of the dynamic puzzle.
\newblock {\em Trends in Cognitive Sciences}, 24(2):135--149, 2020.

\bibitem{kendall1990rank}
M~Kendall and JD~Gibbons.
\newblock {\em Rank correlation methods}.
\newblock Oxford University Press, Oxford, 1990.

\bibitem{khanna2015microstates}
Arjun Khanna, Alvaro Pascual-Leone, Christoph~M Michel, and Faranak Farzan.
\newblock Microstates in resting-state {EEG}: current status and future directions.
\newblock {\em Neuroscience \& Biobehavioral Reviews}, 49:105--113, 2015.

\bibitem{kivela2014multilayer}
Mikko Kivel{\"a}, Alex Arenas, Marc Barthelemy, James~P Gleeson, Yamir Moreno, and Mason~A Porter.
\newblock Multilayer networks.
\newblock {\em Journal of Complex Networks}, 2(3):203--271, 2014.

\bibitem{kong2019spatial}
Ru~Kong, Jingwei Li, Csaba Orban, Mert~R Sabuncu, Hesheng Liu, Alexander Schaefer, Nanbo Sun, Xi-Nian Zuo, Avram~J Holmes, Simon~B Eickhoff, et~al.
\newblock Spatial topography of individual-specific cortical networks predicts human cognition, personality, and emotion.
\newblock {\em Cerebral Cortex}, 29(6):2533--2551, 2019.

\bibitem{korhonen2017consistency}
Onerva Korhonen, Heini Saarim{\"a}ki, Enrico Glerean, Mikko Sams, and Jari Saram{\"a}ki.
\newblock Consistency of {R}egions of {I}nterest as nodes of {fMRI} functional brain networks.
\newblock {\em Network Neuroscience}, 1(3):254--274, 2017.

\bibitem{korhonen2021principles}
Onerva Korhonen, Massimiliano Zanin, and David Papo.
\newblock Principles and open questions in functional brain network reconstruction.
\newblock {\em Human Brain Mapping}, 42(11):3680--3711, 2021.

\bibitem{lin2005functional}
Fa-Hsuan Lin, Teng-Yi Huang, Nan-Kuei Chen, Fu-Nien Wang, Steven~M Stufflebeam, John~W Belliveau, Lawrence~L Wald, and Kenneth~K Kwong.
\newblock Functional {MRI} using regularized parallel imaging acquisition.
\newblock {\em Magnetic Resonance in Medicine}, 54(2):343--353, 2005.

\bibitem{lin2004parallel}
Fa-Hsuan Lin, Kenneth~K Kwong, John~W Belliveau, and Lawrence~L Wald.
\newblock Parallel imaging reconstruction using automatic regularization.
\newblock {\em Magnetic Resonance in Medicine}, 51(3):559--567, 2004.

\bibitem{liu2018co}
Xiao Liu, Nanyin Zhang, Catie Chang, and Jeff~H Duyn.
\newblock Co-activation patterns in resting-state fmri signals.
\newblock {\em Neuroimage}, 180:485--494, 2018.

\bibitem{long2021graph}
Qunfang Long, Suchita Bhinge, Vince~D Calhoun, and T{\"u}lay Adali.
\newblock Graph-theoretical analysis identifies transient spatial states of resting-state dynamic functional network connectivity and reveals dysconnectivity in schizophrenia.
\newblock {\em Journal of Neuroscience Methods}, 350:109039, 2021.

\bibitem{luo2022inside}
Wenjing Luo and R~Todd Constable.
\newblock Inside information: Systematic within-node functional connectivity changes observed across tasks or groups.
\newblock {\em NeuroImage}, 247:118792, 2022.

\bibitem{luo2021within}
Wenjing Luo, Abigail~S Greene, and R~Todd Constable.
\newblock Within node connectivity changes, not simply edge changes, influence graph theory measures in functional connectivity studies of the brain.
\newblock {\em NeuroImage}, 240:118332, 2021.

\bibitem{mehler2018lure}
David Marc~Anton Mehler and Konrad~Paul Kording.
\newblock The lure of misleading causal statements in functional connectivity research.
\newblock {\em arXiv preprint arXiv:1812.03363}, 2018.

\bibitem{mucha2010community}
Peter~J Mucha, Thomas Richardson, Kevin Macon, Mason~A Porter, and Jukka-Pekka Onnela.
\newblock Community structure in time-dependent, multiscale, and multiplex networks.
\newblock {\em Science}, 328(5980):876--878, 2010.

\bibitem{muldoon2016network}
Sarah~Feldt Muldoon and Danielle~S Bassett.
\newblock Network and multilayer network approaches to understanding human brain dynamics.
\newblock {\em Philosophy of Science}, 83(5):710--720, 2016.

\bibitem{pamfil2019relating}
A~Roxana Pamfil, Sam~D Howison, Renaud Lambiotte, and Mason~A Porter.
\newblock Relating modularity maximization and stochastic block models in multilayer networks.
\newblock {\em SIAM Journal on Mathematics of Data Science}, 1(4):667--698, 2019.

\bibitem{pamilo2015correlation}
Siina Pamilo, Sanna Malinen, Jaakko Hotta, and Mika Sepp{\"a}.
\newblock A correlation-based method for extracting subject-specific components and artifacts from group-{fMRI} data.
\newblock {\em European Journal of Neuroscience}, 42(9):2726--2741, 2015.

\bibitem{ryyppo2018regions}
Elisa Ryypp{\"o}, Enrico Glerean, Elvira Brattico, Jari Saram{\"a}ki, and Onerva Korhonen.
\newblock Regions of {I}nterest as nodes of dynamic functional brain networks.
\newblock {\em Network Neuroscience}, 2(4):513--535, 2018.

\bibitem{salehi2020there}
Mehraveh Salehi, Abigail~S Greene, Amin Karbasi, Xilin Shen, Dustin Scheinost, and R~Todd Constable.
\newblock There is no single functional atlas even for a single individual: Functional parcel definitions change with task.
\newblock {\em NeuroImage}, 208:116366, 2020.

\bibitem{salehi2020individualized}
Mehraveh Salehi, Amin Karbasi, Daniel~S Barron, Dustin Scheinost, and R~Todd Constable.
\newblock Individualized functional networks reconfigure with cognitive state.
\newblock {\em NeuroImage}, 206:116233, 2020.

\bibitem{salehi2018exemplar}
Mehraveh Salehi, Amin Karbasi, Xilin Shen, Dustin Scheinost, and R~Todd Constable.
\newblock An exemplar-based approach to individualized parcellation reveals the need for sex specific functional networks.
\newblock {\em NeuroImage}, 170:54--67, 2018.

\bibitem{setsompop2012blipped}
Kawin Setsompop, Borjan~A Gagoski, Jonathan~R Polimeni, Thomas Witzel, Van~J Wedeen, and Lawrence~L Wald.
\newblock Blipped-controlled aliasing in parallel imaging for simultaneous multislice echo planar imaging with reduced g-factor penalty.
\newblock {\em Magnetic Resonance in Medicine}, 67(5):1210--1224, 2012.

\bibitem{smith2011network}
Stephen~M Smith, Karla~L Miller, Gholamreza Salimi-Khorshidi, Matthew Webster, Christian~F Beckmann, Thomas~E Nichols, Joseph~D Ramsey, and Mark~W Woolrich.
\newblock Network modelling methods for {FMRI}.
\newblock {\em NeuroImage}, 54(2):875--891, 2011.

\bibitem{stanley2013defining}
Matthew~L Stanley, Malaak~N Moussa, Brielle~M Paolini, Robert~G Lyday, Jonathan~H Burdette, and Paul~J Laurienti.
\newblock Defining nodes in complex brain networks.
\newblock {\em Frontiers in Computational Neuroscience}, 7:169, 2013.

\bibitem{stanley2016clustering}
Natalie Stanley, Saray Shai, Dane Taylor, and Peter~J Mucha.
\newblock Clustering network layers with the strata multilayer stochastic block model.
\newblock {\em IEEE Transactions on Network Science and Engineering}, 3(2):95--105, 2016.

\bibitem{telesford2017cohesive}
Qawi~K Telesford, Arian Ashourvan, Nicholas~F Wymbs, Scott~T Grafton, Jean~M Vettel, and Danielle~S Bassett.
\newblock Cohesive network reconfiguration accompanies extended training.
\newblock {\em Human Brain Mapping}, 38(9):4744--4759, 2017.

\bibitem{vaiana2018multilayer}
Michael Vaiana and Sarah~Feldt Muldoon.
\newblock Multilayer brain networks.
\newblock {\em Journal of Nonlinear Science}, pages 1--23, 2018.

\bibitem{wang2009parcellation}
Jinhui Wang, Liang Wang, Yufeng Zang, Hong Yang, Hehan Tang, Qiyong Gong, Zhang Chen, Chaozhe Zhu, and Yong He.
\newblock Parcellation-dependent small-world brain functional networks: A resting-state {fMRI} study.
\newblock {\em Human Brain Mapping}, 30(5):1511--1523, 2009.

\bibitem{wu2018approach}
Lei Wu, Arvind Caprihan, Juan Bustillo, Andrew Mayer, and Vince Calhoun.
\newblock An approach to directly link {ICA} and seed-based functional connectivity: Application to schizophrenia.
\newblock {\em NeuroImage}, 179:448--470, 2018.

\bibitem{wu2021tracking}
Lei Wu, Arvind Caprihan, and Vince Calhoun.
\newblock Tracking spatial dynamics of functional connectivity during a task.
\newblock {\em NeuroImage}, 239:118310, 2021.

\bibitem{yu2017selective}
Meichen Yu, Marjolein~MA Engels, Arjan Hillebrand, Elisabeth~CW Van~Straaten, Alida~A Gouw, Charlotte Teunissen, Wiesje~M Van Der~Flier, Philip Scheltens, and Cornelis~J Stam.
\newblock Selective impairment of hippocampus and posterior hub areas in {A}lzheimer’s disease: an {MEG}-based multiplex network study.
\newblock {\em Brain}, 140(5):1466--1485, 2017.

\bibitem{yu2017comparing}
Qingbao Yu, Yuhui Du, Jiayu Chen, Hao He, Jing Sui, Godfrey Pearlson, and Vince~D Calhoun.
\newblock Comparing brain graphs in which nodes are regions of interest or independent components: A simulation study.
\newblock {\em Journal of Neuroscience Methods}, 291:61--68, 2017.

\bibitem{zalesky2012use}
Andrew Zalesky, Alex Fornito, and Ed~Bullmore.
\newblock On the use of correlation as a measure of network connectivity.
\newblock {\em NeuroImage}, 60(4):2096--2106, 2012.

\bibitem{zalesky2010whole}
Andrew Zalesky, Alex Fornito, Ian~H Harding, Luca Cocchi, Murat Y{\"u}cel, Christos Pantelis, and Edward~T Bullmore.
\newblock Whole-brain anatomical networks: does the choice of nodes matter?
\newblock {\em NeuroImage}, 50(3):970--983, 2010.

\bibitem{zang2004regional}
Yufeng Zang, Tianzi Jiang, Yingli Lu, Yong He, and Lixia Tian.
\newblock Regional homogeneity approach to {fMRI} data analysis.
\newblock {\em NeuroImage}, 22(1):394--400, 2004.

\end{thebibliography}

\pagebreak

\begin{center}
 \textbf{\large Supplementary Information for Node-reconfiguring multilayer networks of human brain function}
\end{center}

\setcounter{equation}{0}
\setcounter{figure}{0}
\setcounter{page}{1}
\setcounter{section}{0}
\makeatletter
\renewcommand{\theequation}{S\arabic{equation}}
\renewcommand{\thefigure}{S\arabic{figure}}
\renewcommand{\thesection}{S\arabic{section}}

\setcounter{table}{0}
\renewcommand{\tablename}{Table}
\renewcommand{\thetable}{S\arabic{table}}

\section{Selection of clustering parameters}
\label{sec:SI_selection_of_params}

Defining optimized ROIs requires setting, depending on the approach, one or two free parameters: the percentage of ROIs used for thresholding and the regularization term in the weighted mean consistency approach, the percentage of ROIs used for thresholding in the minimum correlation approach, and the threshold of the NCUT algorithm. To support the selection of these parameter values, we repeated the clustering with a set of parameter values: ROI percentages 10\%, 20\%, 30\%, 40\%, and 50\%, regularization parameter values 10, 100, 250, 500, and 1000 and NCUT threshold values 0.1, 0.2, 0.3, 0.4, and 0.5. This yielded 5 different optimized ROI sets per subject and time window for the minimum correlation and NCUT approaches and 25 different ROI sets per subject and time window for the weighted mean consistency approach. We calculated for each ROI set the weighted mean consistency as

\begin{equation} \label{eq:SI_weighted_consistency}
    \phi_{weighted}(I) = \frac{\sum_I^{ROIs}|I|\phi(I)}{\sum_I^{ROIs}|I|}
\end{equation}

and the size term

\begin{equation} \label{eq:SI_size}
    S(I) = \frac{\sum_I^{ROIs}|I|^2}{(\sum_I^{ROIs}|I|)^2},
\end{equation}

where the summations go over all ROIs $I$ in the ROI set; these two measures correspond to the two terms of equation \ref{eq:weighted_consistency} of the main text. 

\begin{figure}[h!]
    \centering
    \includegraphics[width=\textwidth]{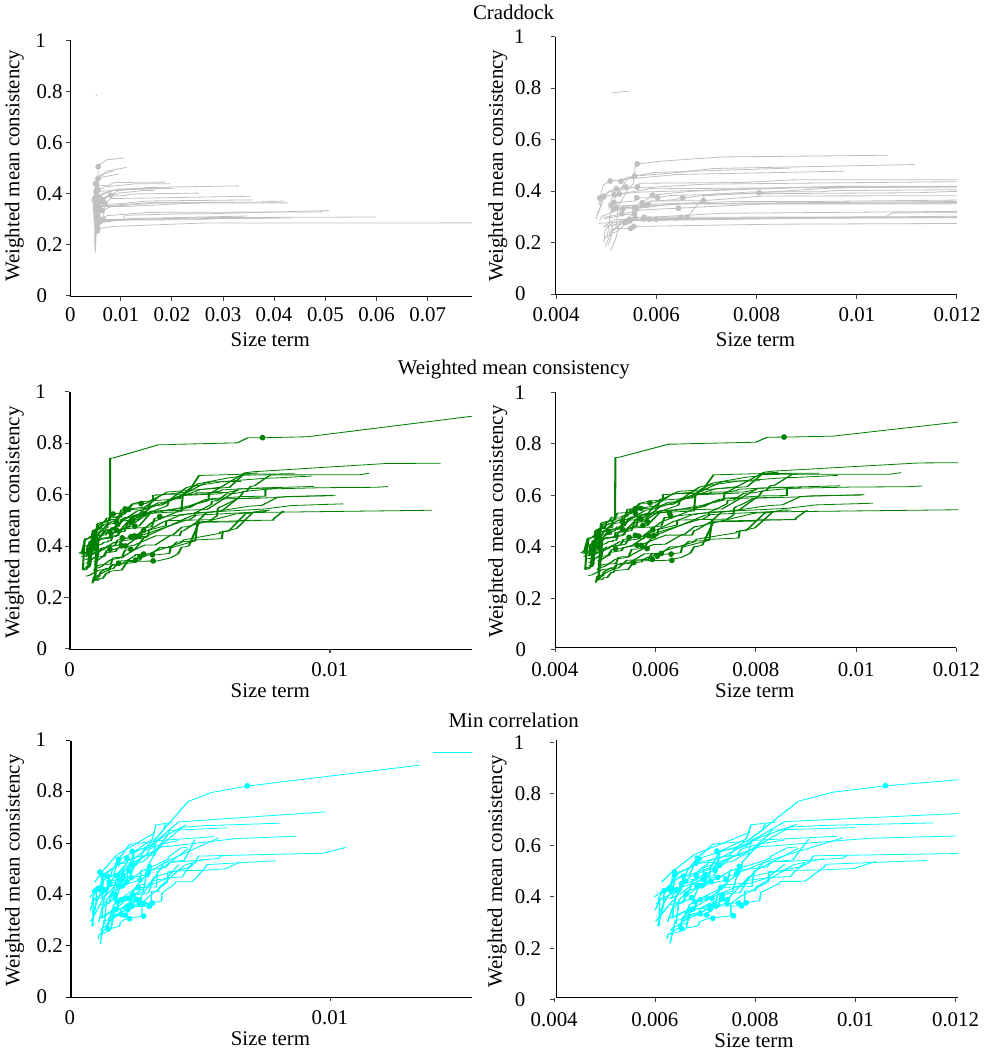}
    \caption{Pareto fronts guide the selection of parameter values. Each line corresponds to the Pareto front calculated for one subject across all time windows for the Craddock (top row), weighted mean consistency (middle row), and minimum correlation (bottom row) parcellation approaches. Markers correspond to parameter values used in the main analysis. Left column shows the full Pareto fronts, while right column shows a zoom-in to the small size term values.}
    \label{fig:SI_pareto_front}
\end{figure}

Together the weighted mean consistency and size term span a space where the best ROI sets are located close to the top left corner, that is, have large weighted mean consistency and a narrow size distribution indicated by a small size term. To find the parameter combinations that produce such ROI sets, we pooled the weighted mean consistency and size term values of each subject across time windows and constructed the Pareto front, or the set of points where the value of the weighted mean consistency cannot be increased without increasing the size term. Each of the points of the Pareto front corresponds to a selection of parameter values that produces an optimal, albeit different, compromise between maximizing the weighted mean consistency and minimizing the size term. The parameter values selected for the main analysis (ROI percentage 30\%, regularization parameter 100, NCUT threshold 0.2) belong to the Pareto front of a vast majority of subjects (Fig. \ref{fig:SI_pareto_front}; Craddock 24/25, Weighted mean cosnistency 23/25, Min correlation 25/25). Note that other parameter combinations belonging to the Pareto front could have further increased the weighted mean consistency of ROIs. However, as the selected values already yielded higher consistency values than the Brainnetome parcellation, we decided to prioritize narrow size distribution 
indicated by the low size term.

{}

\clearpage
\section{Spatial consistency dynamics}

\begin{figure}[h!]
    \centering
    \includegraphics[width=0.7\textwidth]{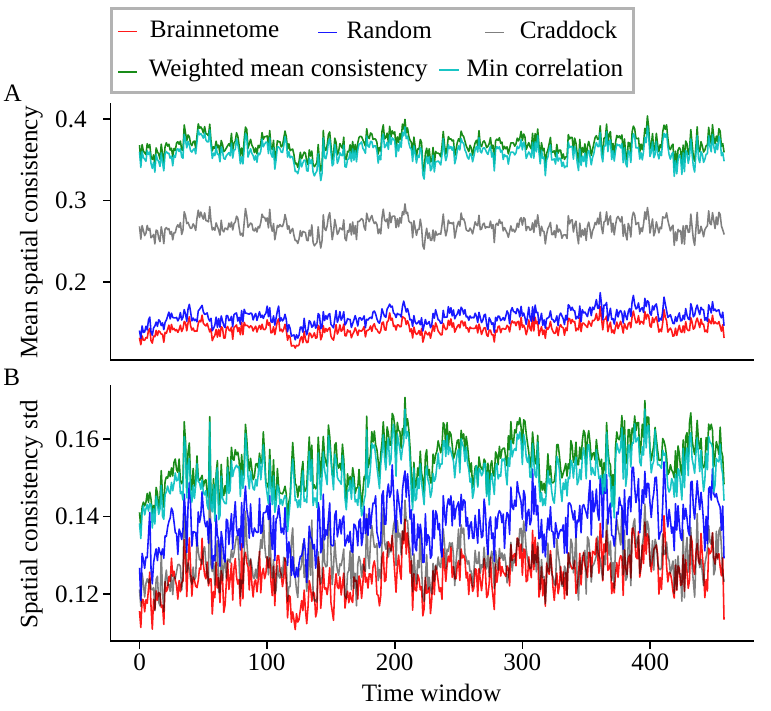}
    \caption{Spatial consistency dynamics. A) Spatial consistency varies in time in all investigated parcellations. However, the changes are relatively small, and no systematic changes or drift are visible in mean spatial consistency calculated across subjects and ROIs. This is not surprising for two reasons. First, systematic, simultaneous changes in multiple subjects are less probable during a naturalistic stimulus then during a  stimulus with block design. Second, the optimized parcellation approaches aim for stable, high spatial consistency and let ROI boundaries change to reach this aim. B) The standard deviation of spatial consistency is relatively high in all parcellations, and does not change systematically in time. For visualization purposes, the imaging runs are attached one after another, and time windows are numbered with a continuous index across runs.}
    \label{fig:SI_consistency_dynamics}
\end{figure}{}

\clearpage
\section{Effect of subcortical areas}

\begin{figure}[h!]
    \centering
    \includegraphics[width=.5\textwidth]{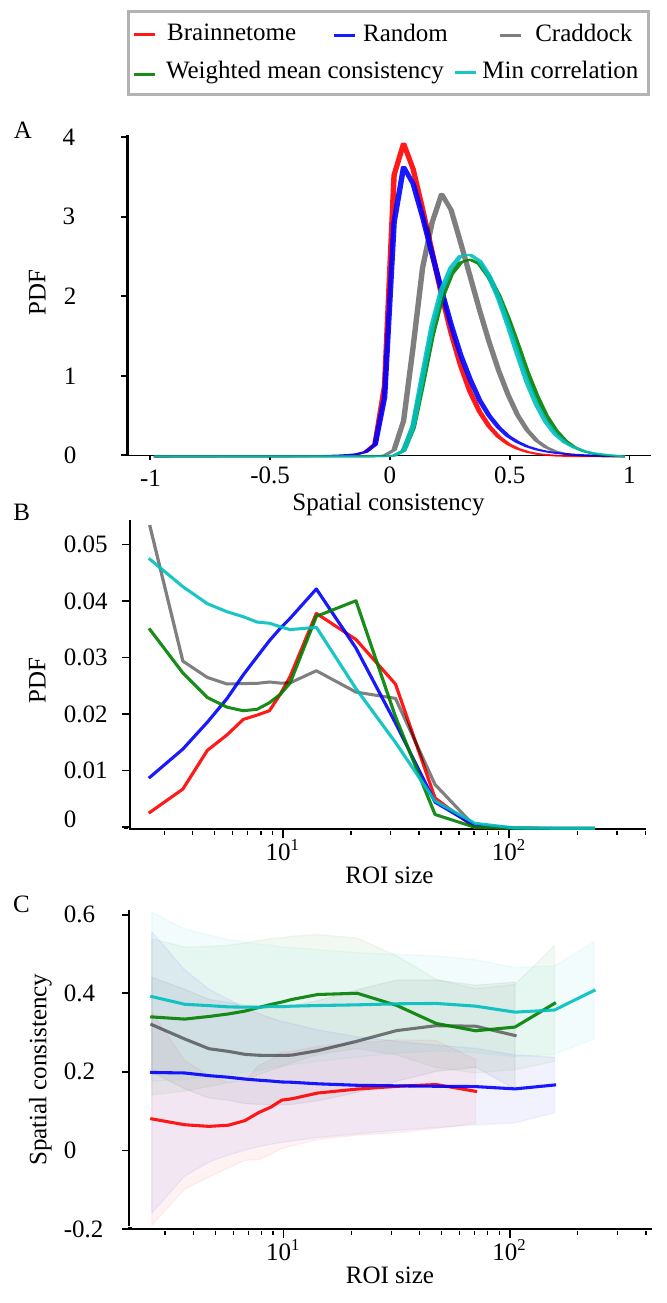}
    \caption{The low spatial consistency of Brainnetome ROIs is not due to subcortical areas. A) Distribution of spatial consistency. B) Distribution of ROI size (defined as the number of voxels in the ROI). C) Average spatial consistency (line) and its standard deviation (shaded area) as a function of ROI size.  All distribution has been calculated over all subjects, runs, time windows, and ROIs, excluding ROIs that contain only a single voxel. In C, spatial consistencies have been binned based on ROI size and then bin-averaged.}
    \label{fig:SI_subcortical_areas}
\end{figure}{}

\clearpage
\section{Spatial consistency with boundaries defined at time $t$}

\begin{figure}[h!]
    \centering
    \includegraphics[width=0.9\textwidth]{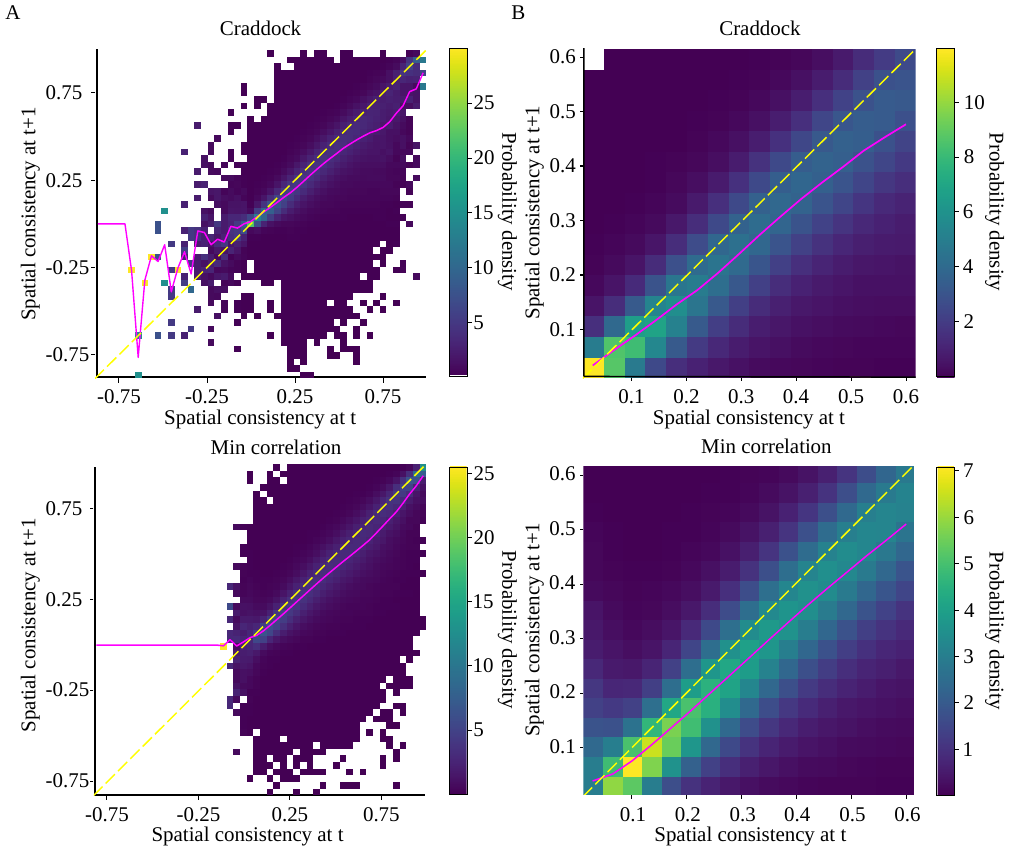}
    \caption{Heatmaps of spatial consistency in time windows t and t + 1 for parcellations produced by the Craddock algorithm and the minimum correlation approach across the whole consistency range (A) and zoomed so that at least 90\% of data is included for all parcellations (B). Consistency is calculated with ROI boundaries defined in window t for both windows, and consistency distribution is calculated for each horizontal bin. Magenta line shows the average spatial consistency in each x bin, and yellow dashed line corresponds to $x = y$. White cells contain no data.}
    \label{fig:SI_consistency_in_next_window}
\end{figure}{}

\clearpage
\section{Amounts of fully stable ROIs}

There are fully stable ROIs in every ROI definition method. The size distributions of these ROIs are presented in Table \ref{table:max_stability_roi_size_distribution}.

\begin{table}[h!]
\centering
\begin{tabular}{l r r r r r}
& \multicolumn{5}{c}{ROI size (voxels)} \\
\cline{2-6}
Parcellation & 1 & 2 & 3 & 4 & 5+ \\
\hline
Brainnetome & 2 582 & 11 691 & 29 595 & 51 948 & 2 671 099 \\
Random & 43 & 35 & 44 & 32 & 82 \\
Craddock & 3 & 6 476 & 2 182 & 1 184 & 3 701 \\
Weighted mean consistency & 82 396 & 15 468 & 8 100 & 4 676 & 9 842 \\
Min correlation & 81 070 & 20 197 & 11 002 & 6 066 & 10 500 \\
\hline
\end{tabular}
\caption{Number of ROIs with stability score exactly 1 (extreme right of Figure 6B in main text) with respect to ROI size.}
\label{table:max_stability_roi_size_distribution}
\end{table}

\end{document}